\documentclass[aps,twocolumn,superscriptaddress,showpacs]{revtex4}
\usepackage{graphicx}% Include figure files

\begin{document}
% You should use BibTeX and apsrev.bst for references
\bibliographystyle{apsrev}

% Use the \preprint command to place your local institutional report
% number on the title page in preprint mode.
% Multiple \preprint commands are allowed.
%\preprint{}

%Title of paper
\title{
Quantum Orders and Spin Liquids in Cs$_2$CuCl$_4$
}
% Optional argument for running titles on pages
%\title[]{}

% repeat the \author .. \affiliation  etc. as needed
% \email, \thanks, \homepage, \altaffiliation all apply to the current
% author. Explanatory text should go in the []'s, actual e-mail
% address or url should go in the {}'s for \email and \homepage.
% Please use the appropriate macro for the type of information

% \affiliation command applies to all authors since the last
% \affiliation command. The \affiliation command should follow the
% other information
% \affiliation can be followed by \email, \homepage, \thanks as well.
\author{Yi Zhou}
\email{yizhou@castu.tsinghua.edu.cn}
\affiliation{Center for Advanced Study, Tsinghua University, Beijing 100084, 
P. R. China}
\author{Xiao-Gang Wen}
%\email{Your e-mail address}
\homepage{http://dao.mit.edu/~wen}
%\thanks{}
%\altaffiliation{}
\affiliation{Department of Physics, Massachusetts Institute of Technology,
Cambridge, Massachusetts 02139}

%Collaboration name if desired (requires use of superscriptaddress
%option in \documentclass). \noaffiliation is required (may also be
%used with the \author command).
%\collaboration can be followed by \email, \homepage, \thanks as well.
%\collaboration{}
%\noaffiliation

\date{\today}

\begin{abstract}
% insert abstract here
Motivated by experiments on Cs$_2$CuCl$_4$ samples, we studied and classified
the symmetric spin liquids on triangular lattice.  We identified 63 $Z_2$ spin
liquids, 30 $U(1)$ spin liquids and 2 $SU(2)$ spin liquids. All those spin
liquids have the same symmetry but different quantum orders. We calculated the
spin spectral functions in some simple spin liquids and compared them to the
one measured by experiments on Cs$_2$CuCl$_4$.  We find that the $U1C\tau
_{+}^0\tau _{-}^0\tau^1$ spin liquid or one of its relatives is consistent
with observed properties of the spin liquid state in Cs$_2$CuCl$_4$. 
We discussed the fine distinctions among those possible spin liquids and the
spin liquids proposed using slave-fermion approach, so that future
experiments can determine which spin liquids actually describe the spin state
in Cs$_2$CuCl$_4$.
\end{abstract}
% insert suggested PACS numbers in braces on next line
\pacs{73.43.Nq,  74.25.-q,  11.15.Ex}
% insert suggested keywords - APS authors don't need to do this
\keywords{Quantum order, Spin liquid, High Tc superconductors, Gauge theory}

%\maketitle must follow title, authors, abstract, \pacs, and \keywords
\maketitle

\section{Introduction}

Our understanding of states of matter and their internal orders has been
dominated by Landau's symmetry breaking theory.\cite{L3726,GL5064} We used to
believe the all possible orders are described by various symmetry breaking
states.  However, after the discovery of fractional quantum Hall (FQH)
states,\cite{TSG8259,L8395} we started to realize that FQH states contain a
new kind of order - topological order\cite{Wtoprev} (which was first proposed
to describe spin liquids found in research of high $T_c$
superconductors\cite{Wtop}). What is new about topological orders is that
topological orders cannot be characterized by symmetry breaking and the
related local order parameters and long range correlations.  Topological
orders are characterized by new set of universal quantum numbers, such as
ground state degeneracy, fractional statistics, edge excitations,
{\it etc}.\cite{Wtoprev}

Recently, a concept of quantum order was proposed\cite{Wqoslpub,Wqogen} to
describe non-symmetry breaking order that generally appear in a quantum state.
The quantum order generalize the topological order to gapless states. To have
a concrete description of quantum order without using order parameters, a new
mathematical object - projective symmetry group (PSG) - was introduced.  The
concept of quantum order and its PSG characterization allow us to understand
quantum phases and quantum phase transitions in a systematic way.  

Just like the group theory allows us to classify symmetry breaking orders, the
PSG characterization of quantum states allows us to classify different quantum
orders\cite{Wqoslpub} and can distinguish two different quantum phases even
when they have exactly the same symmetry.

Also just like the symmetry description of classical order allows us to obtain
low energy properties of system without knowing the details of the systems,
the PSG description of quantum orders also allows us to obtain low energy
properties of system without knowing the details of the systems.  However,
unlike symmetries which produce and protect gapless Nambu-Goldstone
bosons,\cite{N6080,G6154} the quantum orders can produce and protect gapless
collective modes which behave like light and other gapless gauge
bosons.\cite{Wlight,MS0204} Those gapless collective modes can also be
gapless/massless fermions, even when the original theory is purely
bosonic.\cite{KL8842,BFN9833,WZqoind} More recently, it was realized that the
quantum ordered states described by the PSG are actually string-net condensed
states.\cite{Wqoem} The emerging gauge bosons are the fluctuations of
condensed string-nets and emerging fermions are the end of condensed
strings.\cite{LWsta}

In this paper, we would like to further develop the PSG description of quantum
orders. In Ref.\cite{Wqoslpub}, using PSG, a classification of quantum orders
in symmetric spin liquids on 2D square lattice was given.  Here, we would like
expand the results of Ref.\cite{Wqoslpub} to 2D spin liquids which break the
parities $P_x$: $x\to -x$ and $P_y$: $y\to -y$. We will assume the spin
liquids have the following symmetries: two translation symmetries $T_x$,
$T_y$, two parities $P_{xy}$: $(x, y) \to (y,x)$ and $P_{x\bar y}$:
$(x,y)\to (-y,-x)$, and a time reversal symmetry. Such type of 2D spin liquids
was observed recently in Cs$_2$CuCl$_4$ sample.\cite{CTT0135,CTH0203} By
classifying the quantum orders in those spin liquids, we hope to identify the
quantum order in the Cs$_2$CuCl$_4$ sample.

The spin dynamics in Cs$_2$CuCl$_4$ can be described by a 2D spin-1/2 system
on a square lattice with nearest neighbor coupling $J^{\prime }=0.125$meV and
one diagonal coupling $J=0.375$meV in the $\hat{x}+\hat{y}$ direction.  At
temperature $T>T_c=0.62$K$=0.053$meV, the 2D spin system was found to be in a
liquid state. Since $T_c$ is a small energy scale, we will regard the finite
temperature spin liquid state as a zero temperature quantum state. (More
precisely, we assume that we can add additional frustrations to lower $T_c$ to
zero.) The theoretical spin-wave calculations\cite{T9987,MMM9965} and series
expansion calculations\cite{ZMS9967} also suggest the existence of spin
liquids in the above $J$-$J'$ model.  In this paper, we are going concentrate
on the following physical issue: what is this spin liquid in the
Cs$_2$CuCl$_4$ sample. There are several possibilities.

Since $J^{\prime }$ is small, one possibility is that the spin liquid, at low
energies, just behaves like a decoupled 1D spin-1/2 chains. In other words,
the spin liquids is in the same universality class of the decoupled 1D
spin-1/2 chains. It is also possible the spin liquid is not purely 1D and has
intrinsic 2D correlations. The 2D correlation can lead to 2D dispersion for
low lying excitations. Furthermore, there can be several different types of 2D
spin liquids that have the same symmetry as the purely 1D spin liquid.  Since
all those possible spin liquids have the same symmetry, it is difficult to
study them without knowing how to cgaracterize them.

In this paper, using quantum order and its PSG, we construct and characterize
a large class of 2D spin liquids that have the same symmetry as the purely 1D
spin liquid.  In section \ref{sec:ans}, we introduce mean-field ansatz that
describe the symmetric spin liquids - the spin liquids that do not break any
symmetries.  In section \ref{sec:psg} we discuss how to use PSG's to
characterize different mean-field phases (or the universal classes of the
mean-field ansatz).  In section \ref{sec:cls}, we find all the $Z_2$ and
$SU(2)$ PSG's and a large class of $U(1)$ PSG's within the $SU(2)$ slave-boson
approach. This tells us the possible spin liquids that can be constructed use
the $SU(2)$ slave-boson theory.  

Constructing a large class of symmetric spin liquids and obtaining their PSG
characterization are useful in the following sense.  If a spin liquid state is
found in Cs$_2$CuCl$_4$ or some other samples which do not break any symmetry,
then the spin liquid has a good chance to be in the class that we obtained.
Identifying the PSG that characterizes the constructed spin liquid will allow
us to identify many universal properties of the spin liquid. We can check those
universal properties experimentally which allow us to identify the PSGs for
the experimentally observed spin liquids. The classification will also help us
the study the phase transitions between symmetric spin liquids that do not
change any symmetries.

In section \ref{sec:app}, we discuss some simple ansatz that realize some of
the classified mean-field spin liquids.  We calculate the mean-field energies
of the constructed spin liquids to determine which spin liquids are likely to
appear in the Cs$_2$CuCl$_4$ system. We also calculate spin correlation in
those likely spin liquids in section \ref{sec:spe}.  This allows
experimentists to determine which spin liquid actually describe Cs$_2$CuCl$_4$
system (above $T_c$) using neutron scattering and other techniques.  Section
\ref{sec:exp} discusses connection of our results with experiments and
previous theoretical results.

The main point of our calculation is to identify universal properties of
various spin liquid states. This point is highly non-trivial since the
involved spin liquids all have the same symmetry. We show that the crystal
momenta of gapless spin-1 excitations can be used to experimentally
distinguish different spin liquids. 

We would like to pointed out that there are two ways to study possible spin
liquid states in Cs$_2$CuCl$_4$. The first approach is the slave-fermion
approach, where one start with a spin ordered state (such as the spiral state)
and then study the spin liquid induced by strong spin wave
fluctuations.\cite{S9277,CMM0159,CVK0352} One can obtain the spin spectral
function from the slave-fermion approach. The calculated spin spectral
function\cite{CMM0159,CVK0352} agrees well with observed spin spectral
function.\cite{CTT0135,CTH0203} In the  slave-fermion approach, the spin
disordered state (the spin liquid state) always has a gap, with low lying
bosonic excitations.  

In this paper, we are going to use slave-boson approach to study spin liquid
states.  In the slave-boson approach, the spin liquid state can either be
gapped or gapless. The low lying excitations are fermions. In general
slave-boson approach can generate more exotic states then the slave-fermion
approach. We will see later that the spin spectral function obtained from the
slave-boson approach also agrees well with observed spin spectral
function.\cite{CTT0135,CTH0203}

However, the spin liquid obtained in Ref. \cite{S9277,CMM0159} and the spin
liquids obtained in this paper are really different.  For one thing, the spin
liquid obtained in Ref. \cite{S9277,CMM0159} has a gap, while the spin liquids
obtained in this paper is gapless. Because the gap is small, the two spin
liquids behave similarly at finite temperatures.

What is the relation between the slave-boson approach and the slave-fermion
approach?  To understand the relation, we need use the result in Ref.
\cite{Wqoem} where it was shown that the spin liquids obtained from the
slave-boson approach and the slave-fermion approach have a condensation of
nets of closed strings.  The quantum orders in the spin liquids obtained by
both slave-fermion and slave-boson approaches can be described by PSG's.
PSG's simply characterize different string-net condensations.

In the spin liquids obtained by slave-fermion approach, we concentrate on one
type of condensed strings whose ends are bosons (we will call those strings
bosonic strings). The PSG is nothing but the symmetry group of the hopping
Hamiltonian of the ends of the condensed strings.\cite{Wqoem} For different
string condensations, the PSG for the ends of the condensed strings are
different.  Hence PSG can be used to characterize different string
condensations (or different quantum orders).\cite{Wqoem} In slave-fermion
approach, we concentrate on spin liquids with condensation of fermionic
strings. The ends of fermionic strings are fermions.  We can use the PSG for
the ends of fermionic strings to characterize different string condensed (or
quantum ordered) states.  

Some spin liquids contain condensation of both bosonic strings and fermionic
strings. Those states can be characterized by either the PSG for the ends of
the bosonic strings or the PSG for the ends of the fermionic
strings.\cite{Wqoem} Those states can be constructed either by
slave-fermion approach or slave-boson approach.

%In this paper we suggest that the PSG play a role of quantum order in
%symmetric spin liquids and classify the symmetric spin liquids on 2D
%triangular lattices through PSG. As an special example, the nearest neighbor
%spin coupling model is inspected in details.

\section{Symmetric Spin Liquids on a Spin-1/2 System}
\label{sec:ans}

\begin{figure}[htbp]
\begin{center}
\includegraphics[width=2.7in]{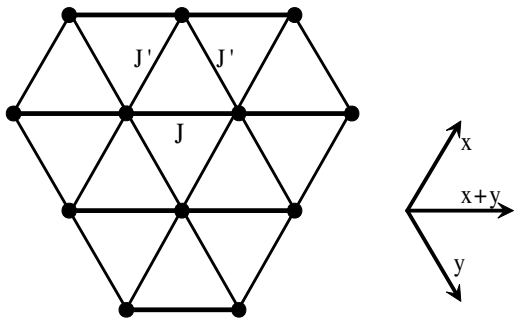}

\end{center}
\caption{The nearest neighbor spin model on a triangular lattice.}
\end{figure}

We consider the spin-1/2 system on a 2D square lattice with nearest-neighbor
coupling $J^{\prime }$ and next-nearest-neighbor coupling $J$ on only one of the
diagonal link $(i_x,i_y)\to (i_x+1,i_y+1)$. Such a lattice can also be
regarded as a triangular lattice which does not has the $60^\circ$ rotation
symmetry. Within the $SU(2)$ slave-boson approach\cite{AZH8845}\cite{DFM8826}, 
general spin wave function can be constructed by introducing a mean-field
Hamiltonian 
\begin{equation}
H_{mean}=-\sum_{\langle i,j\rangle }\left( \psi _i^{\dagger }u_{ij}\psi
_j+h.c.\right) +\sum_ia_0^l\psi _i^{\dagger }\tau ^l\psi _i
\end{equation}
where $\psi ^T=\left( \psi _1,\psi _2\right) $, $u_{ji}^{\dagger }=u_{ij}$, $
\tau ^{1,2,3}$ are the Pauli matrices, and $u_{ij}$ are $2\times 2$ complex
matrices. The collection $\left( u_{ij},a_0^l\tau ^l\right) $ is called a
mean-field ansatz\cite{Wqoslpub}\cite{Wqogen}. For each mean-field ansatz,
we can obtain a mean-field ground state by filling the negative energy
levels of $H_{mean}$ with the spinons $\psi :|\Psi
_{mean}^{(u_{ij},a_0^l\tau ^l)}\rangle $ . The physical spin wave function $
\Psi _{spin}$ can now be obtained by performing a projection (see \cite
{Wqoslpub}): %Gutzwiller projection\cite{G6359}:
\begin{equation}
\Psi_{spin}^{\left( u_{ij},a_0^l\tau ^l\right)}(\{r_i\})=\langle 0|
\prod_{i=1}^{N_{up}}(\psi_{1,r_i}\psi_{2,r_i}) |\Psi_{mean}^{\left(
u_{ij},a_0^l\tau ^l\right) }\rangle
\end{equation}
where $r_i$ is the coordinate of the $i^{th}$ up-spin and $N_{up}$ is the
total number of the up-spin. Here an empty site of $\psi$ represent a
down-spin and a double occupied site represent a up-spin. Single occupied
sties are unphysical states and are projected out.

Our representation does not have explicitly spin rotation symmetry and it is
hard to see spin rotation symmetry from the mean-field ansatz. This problem
is solved in Ref.~\cite{Wqoslpub}. We find that a spin liquid is
spin-rotation symmetric iff $u_{ij}$ satisfies 
\begin{eqnarray}
u_{ij} &=&i\rho _{ij}W_{ij}  \nonumber \\
\rho _{ij} &=&\text{non-negative real number}  \label{rotation} \\
W_{ij} &\in &SU(2)  \nonumber
\end{eqnarray}
In this paper, we will only consider spin liquids with spin rotation
symmetry, and we will assume that the ansatz satisfy the above conditions.

If we regard the ansatz $\left( u_{ij},a_0^l\tau ^l\right)$ as a label of
physical wave function, $\Psi_{spin}^{\left( u_{ij},a_0^l\tau
^l\right)}(\{r_i\})$, then such a label is not a one-to-one label. It has a $
SU(2)$ gauge structure. \textit{i.e.} two ansatz, $\left( u_{ij},a_0^l\tau
^l\right)$ and $\left( W(u_{ij}),W(a_0^l\tau ^l)\right)$, related by a $SU(2)
$ gauge transformation label the same physical wave function: 
\begin{eqnarray}
\Psi _{spin}(\{\alpha _i\}) &=&\langle 0|\prod_if_{i\alpha _i}|\Psi
_{mean}^{(W(u_{ij}),W(a_0^l\tau ^l))}\rangle  \nonumber \\
&=&\langle 0|\prod_if_{i\alpha _i}|\Psi _{mean}^{(u_{ij},a_0^l\tau
^l)}\rangle
\end{eqnarray}
where $W(u_{ij})=W_iu_{ij}W_j^{\dagger }$, $W(a_0^l(i)\tau
^l)=W_ia_0^l(i)\tau ^lW_i^{\dagger }$, and $W_i\in SU(2)$. 
%However, this $SU(2)$ symmetry, describing the ``high energy'' structure, is
%not the issue that we will discuss here. We will discuss the symmetric spin
%liquids described by the effective Hamiltonian (1) which are associated with
%the low energy physics.

\section{Quantum Order and Projective Symmetry Group}
\label{sec:psg}

For the symmetry breaking states, the symmetry is a universal property shared
by all the states in the same phase. Because of this, the symmetry provides a
quantum number that characterize different symmetry breaking orders. Hence we
can say symmetry group (SG) describes the internal orders of symmetry breaking
states.  To characterize and classify different quantum orders which contain
no symmetry breaking, we should construct some universal quantum numbers which
describe different classes of quantum entanglement in the many-body ground
state wave function.  Ref.  \cite{Wqoslpub} and \cite{Wqogen} propose that the
symmetry of the mean-field ansatz $\left( u_{ij},a_0^l\tau ^l\right) $ is a
universal property and serves as a quantum number that characterize the
quantum order in spin liquids. The symmetry of group of the ansatz will be
called the projective symmetry group (PSG). An element of PSG is a combined
operation of a symmetry transformation followed by a gauge transformation. By
definition, a PSG is formed by all the combined operations that leaves the
ansatz unchanged.

Because of the $SU(2)$ gauge structure, the $\left(u_{ij}, a_0^l\tau^l\right)
$ labeling of the physical spin wave function $\Psi _{spin}(\{\alpha _i\})$
is not a one-to-one labeling. Two mean-field ansatz differed by a $SU(2)$
gauge transformation give rise to the same spin wave function. Because of
this, it is a non-trivial task to find out the symmetry of a spin liquid
label by an ansatz $\left(u_{ij}, a_0^l\tau^l\right)$. In order for a spin
wave function to have a symmetry, its corresponding ansatz is only required
to be invariant under the symmetric transformation followed by a proper $
SU(2)$ gauge transformation. Thus given two spin wave functions with the
same symmetry, their ansatz can be invariant under the same symmetric
transformations followed by \emph{different} gauge transformations. In this
case, the two spin liquids with the same symmetry can have different PSG's.
We see that the PSG characterization is more refined than the symmetry group
characterization. PSG contains some information about the phase of spin wave
function.

To understand the precise relation between PSG and SG, we need to introduce a
subgroup of PSG - the invariant gauge group (IGG).\cite{Wqoslpub}\cite{Wqogen}
IGG is a special subgroup of PSG, which is formed by pure gauge
transformations that leave the ansatz unchanged, 
\begin{equation}
IGG\equiv \left\{ W_i|W_iu_{ij}W_j^{\dagger }=u_{ij},W_i\in SU(2)\right\}
\label{IGG}
\end{equation}
Then SG and PSG are related by $SG=PSG/IGG$.

Spin liquids may support a special kind of low energy collective excitations
- gauge fluctuations.\cite{BA8880} It was shown that\cite{Wsrvb,Wqoslpub}
the gauge group of those low energy gauge fluctuations is nothing but the
IGG of the corresponding ansatz. We see that IGG of an ansatz is very
important. In this paper, we will consider only three kinds of IGG, $SU(2)$, 
$U(1)$ and $Z_2$. Correspondingly, we call the corresponding spin liquids $
SU(2)$, $U(1)$ and $Z_2$ spin liquids.

Mathematically, PSG is defined as, 
\begin{equation}
PSG\equiv \{G_U|G_UU(u_{ij})=u_{ij},G_U(i)\in SU(2)\}
\end{equation}
where $U(u_{ij})=\tilde{u}_{ij}\equiv u_{U(i),U(j)}$, $G_UU(\tilde{u}
_{ij})\equiv G_U(i)\tilde{u}_{ij}G_U^{\dagger }(j)$, $U$ generates the
symmetry transformation and $G_U$ is the associated gauge transformation.

A PSG may change under a gauge transformation $W$. From $
WG_U(u_{ij})=W(u_{ij})$, where $W(u_{ij})\equiv W_iu_{ij}W_j^{\dagger }$, we
find that $WG_UUW^{-1}W(u_{ij})=W(u_{ij})$. Therefore if $G_UU$ is in the PSG
of ansatz $u_{ij}$, then $WG_UUW^{-1}$ is in the PSG of the gauge transformed
ansatz $W(u_{ij})$. We see that the gauge transformation $G_U$ associated with
the transformation $U$ changes in the following way 
\begin{equation}
G_U(i)\rightarrow W(i)G_U(i)W(U(i))^{\dagger }
\end{equation}
under a $SU(2)$ gauge transformation.

Since PSG is a property of an ansatz, we can group all ansatz sharing the
same PSG together to form a class. Such a class is the universal class of
quantum states that corresponds a quantum phase.

Now let us consider some simple examples of spin liquids described by the
ansatz $\left( u_{ij},a_0^l\tau ^l\right) $. The first example is 
\begin{equation}
u_{i,i+m}=u_m=u_m^l\tau ^l
\end{equation}
where $u_m^l$ are real, $l=1,2,3$. It is easy to obtain the spinon
dispersion of such an ansatz, 
\[
E_{\pm }(k)=\pm \sqrt{\sum_l(u_k^l-a_0^l)^2} 
\]
where $u_k^l=\sum_mu_m^le^{ik\cdot m}$. The Brillouin zone is $k_x,k_y\in
(-\pi ,\pi )$.

The second example is 
\begin{eqnarray}
u_{i,i+\hat{x}} &=&i\chi \tau ^0+\eta \tau ^3,  \nonumber \\
u_{i,i+\hat{y}} &=&\left( -\right) ^{i_x}\left( i\chi \tau ^0+\eta \tau
^3\right) , \\
u_{i,i+\hat{x}+\hat{y}} &=&\left( -\right) ^{i_x}\lambda \tau ^1,  \nonumber
\end{eqnarray}
where $\chi ,\eta $ and $\lambda $ are real. Its spinon spectrum is
determined by 
\begin{eqnarray*}
H(k) &=&\left( \chi \sin k_x\tau ^0+\eta \cos k_x\tau ^3\right) \otimes \tau
^1 \\
&&+\left( \chi \sin k_y\tau ^0+\eta \cos k_y\tau ^3\right) \otimes \tau ^3 \\
&&+\lambda \tau ^1\otimes \tau ^2\sin (k_x+k_y)+a_0^l\tau ^l\otimes \tau ^0
\end{eqnarray*}
where the Brillouin zone is $k_x\in (-\pi /2,\pi /2),k_y\in (-\pi ,\pi )$.
The four bands of spinon dispersion have a form of $\pm E_1(k),\pm E_2(k)$.
It seems strange to find that the spinon spectrum is defined only on half of
the lattice Brillouin zone. However, this is not inconsistent with the
translation symmetry since the single spinon excitation is not physical.
Only two-spinon excitations correspond to physical excitations and their
spectrum should be defined on the full Brillouin zone. The two-spinon
spectrum defined on the full Brillouin zone can be constructed form
single-spinon spectrum\cite{Wqoslpub} 
\begin{eqnarray}
E_{2s}(k) &=&E_{\alpha _1}(k_1)+E_{\alpha _2}(k_2)  \nonumber \\
k &=&k_1+k_2+n\pi \hat{x}
\end{eqnarray}
where $n=0,1$ and $\alpha _{1,2}=1,2$ is the sub-index of the single spinon
dispersion $\pm E_1(k),\pm E_2(k)$. We note that the physical spin-1
excitations are formed by two-spinon excitations in the mean-field theory.
Thus the The two-spinon spectrum $E_{2s}$ is also the spectrum of physical
spin-1 excitation and can be measured in experiments.

\section{Classification of Symmetric Spin Liquids}
\label{sec:cls}

In this section, we classify symmetric spin liquids on triangular lattices.
We consider the spin liquids which are invariant under translation
transformation $T_x(i\rightarrow i+\hat{x})$ and $T_y(i\rightarrow i+\hat{y}
) $, parity transformation $P_{xy}\left( (i_x,i_y)\rightarrow
(i_y,i_x)\right) $ and $P_{x\bar{y}}\left( (i_x,i_y)\rightarrow
(-i_y,-i_x)\right) $, spin rotation transformation, spin-parity
transformation $T^{*}\left( S_x\rightarrow S_x,S_y\rightarrow
-S_y,S_z\rightarrow S_z\right) $. The spin rotation symmetry requires the
ansatz take the form of Eq.(\ref{rotation}). Under the spin-parity
transformation $T^{*}$, the three components of spin change as $
S_x\rightarrow S_x,$\ $S_y\rightarrow -S_y,$\ $S_z\rightarrow S_z$. This
transformation can be realized through wave-function transformation $\Phi
\rightarrow \Phi ^{*}$, 
\[
\langle \vec{S}\rangle =\frac 12\int \Phi ^{*}f^{\dagger }\vec{\sigma}f\Phi
\rightarrow \frac 12\int \Phi f^{\dagger }\vec{\sigma}f\Phi ^{*}, 
\]
under the $S_z$\ representation $f$\ and $f^{\dagger }$\ are real, then 
\begin{eqnarray*}
\langle \vec{S}\rangle &=&\langle \vec{S}\rangle ^{*}=\frac 12\int \Phi
^{*}f^{\dagger }\vec{\sigma}f\Phi \\
&\rightarrow &\frac 12\int \Phi f^{\dagger }\vec{\sigma}f\Phi ^{*}=\left(
\frac 12\int \Phi f^{\dagger }\vec{\sigma}f\Phi ^{*}\right) ^{*} \\
&=&\frac 12\int \Phi ^{*}f^{\dagger }\vec{\sigma}^{*}f\Phi ,
\end{eqnarray*}
which leads to $S_x\rightarrow S_x,$\ $S_y\rightarrow -S_y,$\ $
S_z\rightarrow S_z$. The ansatz $u_{ij}$\ transfer as $u_{ij}\rightarrow
u_{ij}^{*}$, combined with a gauge transformation $u_{ij}\rightarrow \left(
i\tau ^2\right) u_{ij}\left( -i\tau ^2\right) $, we denote this combined
transformation as\ 
\begin{equation}
T^{*}:u_{ij}\rightarrow -u_{ij}.  \label{T*}
\end{equation}

%As an example, the lattice in Fig.1 offer such a example. We have
%found that there are three kinds of spin liquids, say, $SU(2)$,
%$U(1)$ and $Z_2$ spin liquids in the last section. In the
%following we will classify these three kinds spin liquids through
%PSG's respectively.

Let us firstly discuss the method to classify the PSG briefly and leave the
details to the appendix. For two given symmetry transformations, whose
corresponding elements in a PSG, must satisfy some algebraic relations.
Solving these equations allows us to construct a PSG which will be called
the algebraic PSG. The new name algebraic PSG is introduced to distinguish
them from the invariant PSG defined in the previous section. Any invariant
PSG will be algebraic PSG. But an algebraic PSG may not be an invariant PSG
unless there does exist an ansatz such that the algebraic PSG is the total
symmetry group of the ansatz.

As an example, let us consider the two translations $T_x$ and $T_y$, which
satisfy the following relation 
\begin{equation}
T_xT_yT_x^{-1}T_y^{-1}=1.
\end{equation}
From the definition of PSG, we find that the two elements of PSG, $G_xT_x$
and $G_yT_y$, must satisfy 
\begin{eqnarray}
&&G_xT_xG_yT_y(G_xT_x)^{-1}(G_yT_y)^{-1}  \nonumber \\
&=&G_xT_xG_yT_yT_x^{-1}G_x^{-1}T_y^{-1}G_y^{-1}  \nonumber \\
&=&G_x\left( i\right) G_y\left( i-\hat{x}\right) G_x^{-1}\left( i-\hat{y}
\right) G_y^{-1}\left( i\right) \in \mathcal{G}.  \label{TxTy}
\end{eqnarray}
Hereafter we will denote the IGG as $\mathcal{G}$. Each solution of the
equation (\ref{TxTy}) corresponds to an algebraic PSG for $T_x$ and $T_y$.

By adding other symmetry transformations, we can find and classify all the
algebraic PSG associated with a given symmetry group. Since an invariant PSG
is always an algebraic PSG, we can check whether the algebraic PSG is an
invariant PSG through constructing an explicit ansatz $u_{ij}$. If an
algebraic PSG supports an ansatz $u_{ij}$ with no addition symmetry, then it
is an invariant PSG. Through this method, we can classify the symmetric spin
liquids through PSG.

In the appendix, we classify spin liquids with the above mentioned symmetries
through PSG's. We limit ourselves to $SU(2)$, $U(1)$ and $Z_2$ spin liquids
(we only consider those PSG's whose IGG is one of $SU(2)$, $U(1)$ and $Z_2$
). We found that there are 63 kinds of $Z_2$ spin liquids, 30 kinds of $U(1)$
spin liquids and 2 kinds of $SU(2)$ spin liquids with $T_{x,y}$, 
$P_{xy,x\bar{y}}$, $T^{*}$ and spin rotation symmetries.

\section{Application to Nearest-Neighbor Spin Coupling Model}
\label{sec:app}

In this section, we present our classification results via some simple
examples. We will assume only $u_{i,i+x}$, $u_{i,i+x}$, and $u_{i,i+x+y}$
are no zero. 
%$J_{\hat{x}}=J_{\hat{y}}=J^{\prime },J_{\hat{x}+\hat{y}}=J$ and $J_{ij}=0$ for other
%bonds. The lattice is described by Fig. 1.
In this case, we have the following 7 kinds of $Z_2$ spin liquids and 3
kinds of $U(1)$ (Other $Z_2$ and $U(1)$ spin liquids require non-vanishing $
u_{ij}$ on longer bonds.) 
%spin liquids support $J^{\prime }\neq 0$ and $J\neq 0$ in this model.

The 7 $Z_2$ spin liquids are:\newline
(1) $Z2A\tau ^0\tau _{+}^0\tau _{+}^3$ spin liquid 
\begin{eqnarray}
u_{i,i+\hat{x}} &=&\chi \tau ^1+\eta \tau ^2,  \nonumber \\
u_{i,i+\hat{y}} &=&\chi \tau ^1+\eta \tau ^2,  \label{Ex1} \\
u_{i,i+\hat{x}+\hat{y}} &=&\lambda \tau ^1,  \nonumber \\
a_0^1 &=&a_1,a_0^2=a_2,a_0^3=0  \nonumber
\end{eqnarray}
(2) $Z2A\tau ^1\tau _{+}^1\tau _{+}^3$ spin liquid 
\begin{eqnarray}
u_{i,i+\hat{x}} &=&\chi \tau ^1+\eta \tau ^2,  \nonumber \\
u_{i,i+\hat{y}} &=&\chi \tau ^1-\eta \tau ^2,  \label{Ex2} \\
u_{i,i+\hat{x}+\hat{y}} &=&\lambda \tau ^1,  \nonumber \\
a_0^1 &=&a_1,a_0^{2,3}=0  \nonumber
\end{eqnarray}
(3) $Z2A\tau ^0\tau _{-}^1\tau _{-}^3$ spin liquid 
\begin{eqnarray}
u_{i,i+\hat{x}} &=&i\chi \tau ^0+\eta \tau ^3,  \nonumber \\
u_{i,i+\hat{y}} &=&i\chi \tau ^0+\eta \tau ^3, \\
u_{i,i+\hat{x}+\hat{y}} &=&\lambda \tau ^1,  \nonumber \\
a_0^1 &=&a_1,a_0^{2,3}=0  \nonumber
\end{eqnarray}
(4) $Z2A\tau ^1\tau _{-}^3\tau _{-}^3$ spin liquid 
\begin{eqnarray}
u_{i,i+\hat{x}} &=&i\chi \tau ^0+\eta \tau ^3,  \nonumber \\
u_{i,i+\hat{y}} &=&i\chi \tau ^0-\eta \tau ^3, \\
u_{i,i+\hat{x}+\hat{y}} &=&\lambda \tau ^1,  \nonumber \\
a_0^1 &=&a_1,a_0^{2,3}=0  \nonumber
\end{eqnarray}
(5) $Z2B\tau ^3\tau _{-}^0\tau _{+}^3$ spin liquid 
\begin{eqnarray}
u_{i,i+\hat{x}} &=&\chi \tau ^1+\eta \tau ^2,  \nonumber \\
u_{i,i+\hat{y}} &=&\left( -\right) ^{i_x}\left( -\chi \tau ^1-\eta \tau
^2\right) , \\
u_{i,i+\hat{x}+\hat{y}} &=&\left( -\right) ^{i_x}\lambda \tau ^1,  \nonumber
\\
a_0^{1,2,3} &=&0  \nonumber
\end{eqnarray}
(6) $Z2B\tau ^1\tau _{-}^2\tau _{+}^3$ spin liquid 
\begin{eqnarray}
u_{i,i+\hat{x}} &=&\chi \tau ^1+\eta \tau ^2,  \nonumber \\
u_{i,i+\hat{y}} &=&\left( -\right) ^{i_x}\left( \chi \tau ^1-\eta \tau
^2\right) , \\
u_{i,i+\hat{x}+\hat{y}} &=&\left( -\right) ^{i_x}\lambda \tau ^2,  \nonumber
\\
a_0^{1,2,3} &=&0  \nonumber
\end{eqnarray}
(7) $Z2B\tau ^3\tau _{-}^1\tau _{-}^3$ spin liquid 
\begin{eqnarray}
u_{i,i+\hat{x}} &=&i\chi \tau ^0+\eta \tau ^3,  \nonumber \\
u_{i,i+\hat{y}} &=&\left( -\right) ^{i_x}\left( i\chi \tau ^0+\eta \tau
^3\right) , \\
u_{i,i+\hat{x}+\hat{y}} &=&\left( -\right) ^{i_x}\lambda \tau ^1,  \nonumber
\\
a_0^{1,2,3} &=& 0.  \nonumber
\end{eqnarray}

The 3 $U(1)$ spin liquids are:\newline
(1) $U1A\tau ^0\tau _{+}^0\tau _{+}^1$ spin liquids: 
\begin{eqnarray}
u_{i,i+\hat{x}} &=&\chi \tau ^3,  \nonumber \\
u_{i,i+\hat{y}} &=&\chi \tau ^3, \\
u_{i,i+\hat{x}+\hat{y}} &=&\lambda \tau ^3,  \nonumber \\
a_0^{1,2} &=&0,a_0^3=a_3.  \nonumber
\end{eqnarray}
(2) $U1B\tau ^1\tau _{-}^0\tau _{+}^1$ spin liquids: 
\begin{eqnarray}
u_{i,i+\hat{x}} &=&\chi \tau ^3,  \nonumber \\
u_{i,i+\hat{y}} &=&-\left( -\right) ^{i_x}\chi \tau ^3,  \label{Bphase} \\
u_{i,i+\hat{x}+\hat{y}} &=&\left( -\right) ^{i_x}\lambda \tau ^3,  \nonumber
\\
a_0^{1,2,3} &=&0.  \nonumber
\end{eqnarray}
(3) $U1C\tau _{+}^0\tau _{-}^0\tau ^1$ spin liquids: 
\begin{eqnarray}
u_{i,i+\hat{x}} &=&\chi \tau ^2,  \nonumber \\
u_{i,i+\hat{y}} &=&-\chi \tau ^2,  \label{Cphase} \\
u_{i,i+\hat{x}+\hat{y}} &=&\lambda \tau ^3,  \nonumber \\
a_0^{1,2} &=&0,a_0^3=a_3.  \nonumber
\end{eqnarray}
The labeling scheme of these spin liquids is defined in the appendix.

We have performed the self-consistent mean-field calculation for our 
$J$-$J^{\prime }$ model. We found many self-consistent mean-field solutions. The energies
of four of them are plotted in Fig. 2, where we have set $J+J^{\prime }=1$. To
understand the physical properties of those mean-field states, we need to
find their PSGs.

% support the
%following 3 kinds of $U(1)$ spin liquids and $Z2A\tau ^1\tau
%_{+}^1\tau _{+}^3$ spin liquid. Fig. 2. show the mean-field energy
%for various phases in the nearest-neighbor spin coupling model,
%where we have set $J+J^{\prime }=1$.

\begin{figure}[htbp]
\begin{center}
\includegraphics[width=3.6in]{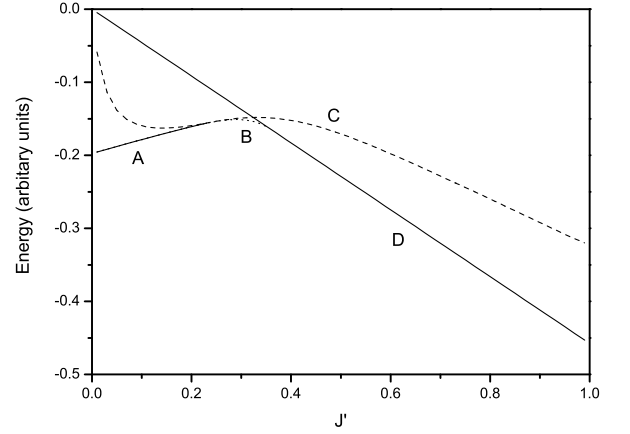}

\end{center}
\caption{The mean-field energies for various phases in the nearest neighbor
spin coupling system. (A) the $SU(2)\times SU(2)$ spin liquids in Eq.(\ref
{Aphase}). (B) the $U1B\tau^1\tau _{-}^0\tau _{+}^1$ state in Eq.(\ref
{Bphase}). (C) the $U1C\tau _{+}^0\tau _{-}^0\tau ^1$ state in Eq.(\ref
{Cphase}). (D) the $\pi$-flux phase ($SU2B$ state) in Eq.(\ref{Dphase}).}
\end{figure}

The A phase in Fig. 2 is a one-dimensional state. Its IGG is $SU^\infty(2)$,
one $SU(2)$ for each decoupled 1D chain. We will call such a state $
SU^\infty(2)$ spin liquid. Its ansatz is given by 
\begin{equation}
u_{i,i+\hat{x}+\hat{y}}=\lambda \tau ^3,a_0^{1,2,3}=0,  \label{Aphase}
\end{equation}
with spinon dispersion 
\begin{equation}
E(k)=\pm 2\lambda \cos \left( k_x+k_y\right)
\end{equation}
This phase, favored by small $J^{\prime }$, corresponds to the phase of decoupled 1D
spin chain mentioned earlier.

The B phase is the $U1B\tau ^1\tau _{-}^0\tau _{+}^1$ spin liquid. Its four
spinon bands are given by 
\begin{equation}
\pm 2\sqrt{\chi ^2(\cos ^2k_x+\cos ^2k_y)+\lambda ^2\sin ^2(k_x+k_y)}
\end{equation}
The spinon dispersion min($E_1(k),E_2(k)$) and the lower edge of the
spectrum $E_{2s}(k)$ of the physical spin-1 excitations are plotted as
functions of $(k_x/\pi ,k_y/\pi )$ in Fig. 3. This phase is one of 2D spin
liquid phases.

\begin{figure}[htbp]
\begin{center}
\includegraphics[width=3.0in]{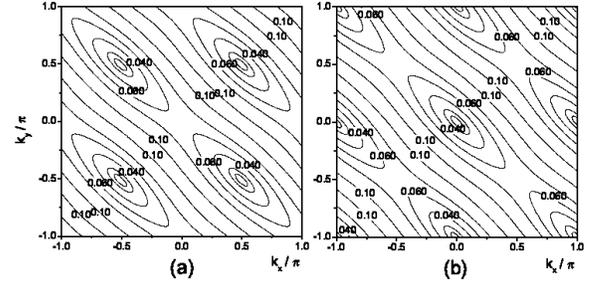}

\end{center}
\caption{(a) Contour plot of the spinon dispersion $E_{+}(k)$ as a function
of $(k_x/\pi ,k_y/\pi )$ for the $U1B\tau ^1\tau _{-}^0\tau_{+}^1$ state.
(b) Lower edge of the two-spinon spectrum $E_{2s}(k)$ as a function of $
(k_x/\pi ,k_y/\pi )$ for the $U1B\tau ^1\tau_{-}^0\tau _{+}^1$ state.}
\end{figure}

The C phase has a PSG $U1C\tau _{+}^0\tau _{-}^0\tau ^1$. The spinon
dispersion is given by 
\begin{eqnarray}
E_{+}(k) &=&2\sqrt{\left[ \lambda \cos \left( k_x+k_y\right) -a_3\right]
^2+\chi ^2\left( \cos k_x-\cos k_y\right) ^2}  \nonumber \\
&&
\end{eqnarray}
The spinon dispersion $E_{+}(k)$ and the lower edge of the spin-1 spectrum $
E_{2s}(k)$ are plotted as functions of $(k_x/\pi ,k_y/\pi )$ in Fig. 4. This
phase is another 2D spin liquid phase.

\begin{figure}[htbp]
\begin{center}
\includegraphics[width=3.0in]{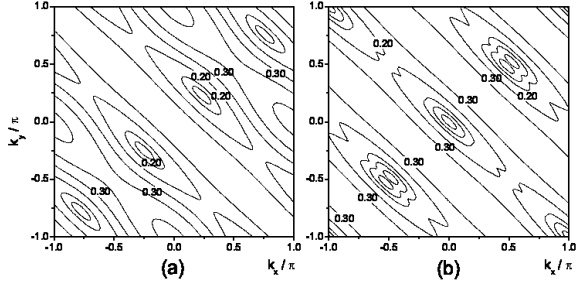}

\end{center}
\caption{(a) Contour plot of the spinon dispersion $E_{+}(k)$ as a function
of $(k_x/\pi ,k_y/\pi )$ for the $U1C\tau _{+}^0 \tau _{-}^0\tau ^1$ state.
(b) Lower edge of the two-spinon spectrum $E_{2s}(k)$ as a function of $
(k_x/\pi ,k_y/\pi )$ for the $U1C\tau _{+}^0\tau _{-}^0\tau ^1$ state.}
\end{figure}

The D phase in Fig.2 is the $\pi $-flux phase ($SU2B$ spin liquid), which is
given by 
\begin{eqnarray}
u_{i,i+\hat{x}} &=&\chi \tau ^1,  \nonumber \\
u_{i,i+\hat{y}} &=&\chi \tau ^2,  \label{Dphase} \\
a_0^{1,2,3} &=&0.  \nonumber
\end{eqnarray}
The spinon dispersion is 
\begin{equation}
E_{+}(k)=2\chi \sqrt{\cos ^2k_x+\cos ^2k_y}
\end{equation}
The spinon dispersion $E_{+}(k)$ and the lower edge of the spin-1 spectrum $
E_{2s}(k)$ are plotted as functions of $(k_x/\pi ,k_y/\pi )$ in Fig. 5. 
\begin{figure}[tbph]
\begin{center}
\includegraphics[width=3.0in]{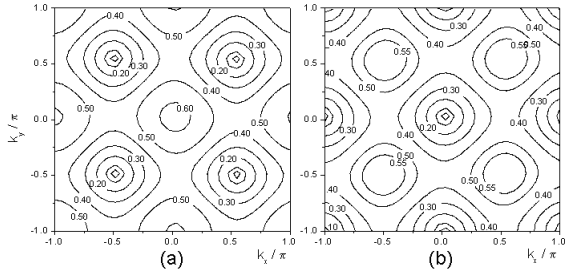}

\end{center}
\caption{a) Contour plot of the spinon dispersion $E_{+}(k)$ as a function
of $(k_x/\pi ,k_y/\pi )$ for the $\pi $-flux state. (b) Lower edge of the
two-spinon spectrum $E_{2s}(k)$ as a function of $(k_x/\pi ,k_y/\pi )$ for
the $\pi $-flux state.}
\end{figure}

Now we consider two other $Z_2$ spin liquids. The first one is the $Z2A\tau
^0\tau _{+}^0\tau _{+}^3$ spin liquid state in Eq.(\ref{Ex1}). The energy
dispersion of the $Z2A\tau ^0\tau _{+}^0\tau _{+}^3$ state is given by 
\begin{eqnarray}
E_{+}(k) &=&2\sqrt{\epsilon _1^2+\epsilon _2^2}  \nonumber \\
\epsilon _1 &=&\chi \left( \cos k_x+\cos k_y\right) +\lambda \cos \left(
k_x+k_y\right) -a_1  \nonumber \\
\epsilon _2 &=&\eta \left( \cos k_x+\cos k_y\right) 
\end{eqnarray}
The spinon dispersion $E_{+}(k)$ and the lower edge of the spin-1 spectrum $
E_{2s}(k)$ are plotted as functions of $(k_x/\pi ,k_y/\pi )$ in Fig. 6.

\begin{figure}[htbp]
\begin{center}
\includegraphics[width=3.2in]{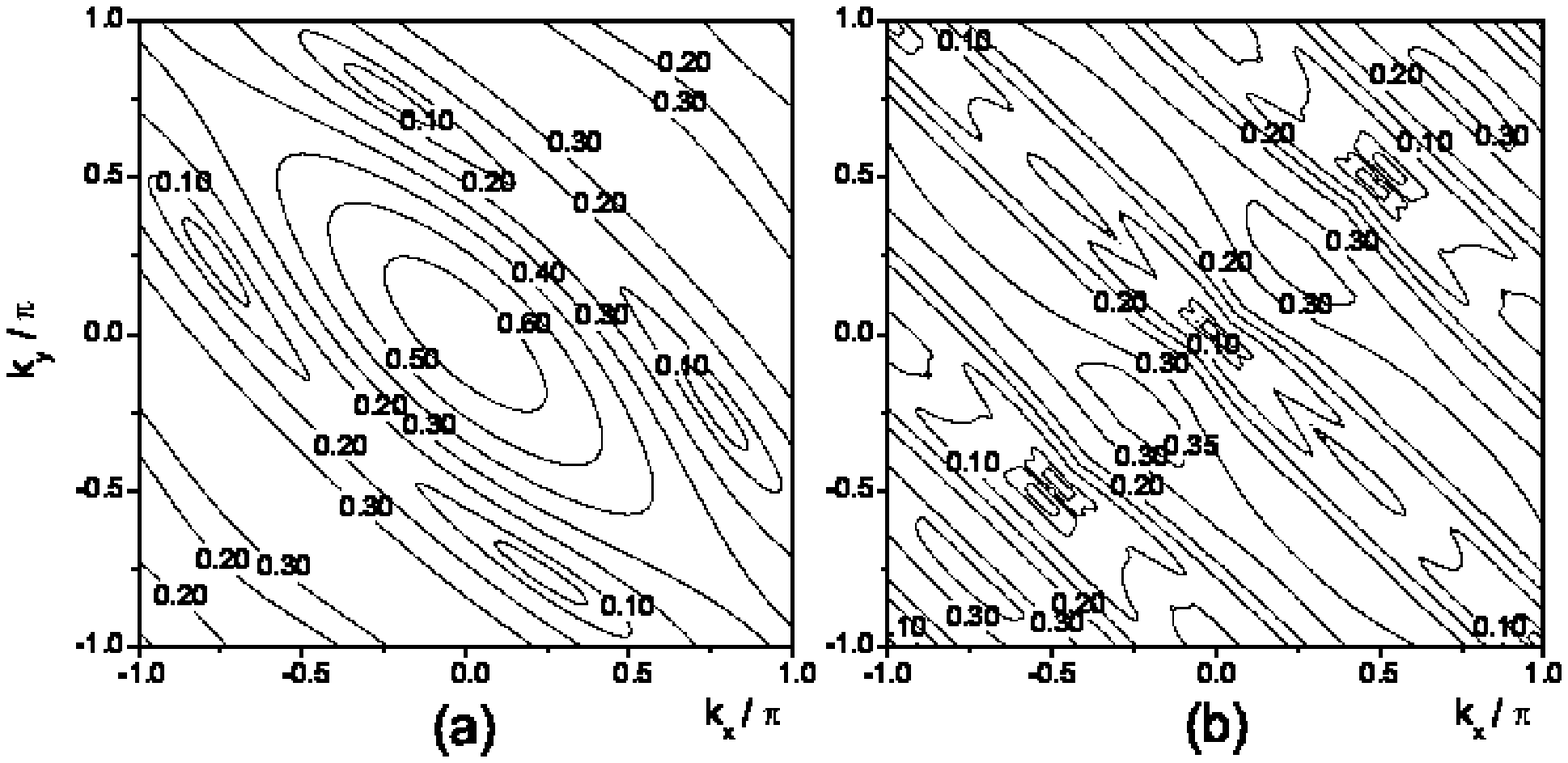}

\end{center}
\caption{((a) Contour plot of the spinon dispersion $E_{+}(k)$ as a function
of $(k_x/\pi ,k_y/\pi )$ for the $Z2A\tau ^0\tau _{+}^0\tau _{+}^3$ state.
(b) Lower edge of the two-spinon spectrum $E_{2s}(k)$ as a function of $
(k_x/\pi ,k_y/\pi )$ for the $Z2A\tau ^0\tau _{+}^0\tau _{+}^3$ state.}
\end{figure}

The second one is the $Z2A\tau ^0\tau _{+}^0\tau _{+}^3$ spin liquid state
in Eq.(\ref{Ex2}). Energy dispersion of the $Z2A\tau ^0\tau _{+}^0\tau _{+}^3
$ state is given by 
\begin{eqnarray}
E_{+}(k) &=&2\sqrt{\epsilon _1^2+\epsilon _2^2}  \nonumber \\
\epsilon _1 &=&\chi \left( \cos k_x+\cos k_y\right) +\lambda \cos \left(
k_x+k_y\right) -a_1  \nonumber \\
\epsilon _2 &=&\eta \left( \cos k_x-\cos k_y\right) 
\end{eqnarray}
The spinon dispersion $E_{+}(k)$ and the lower edge of the spin-1 spectrum $
E_{2s}(k)$ are plotted as functions of $(k_x/\pi ,k_y/\pi )$ in Fig. 7.

\begin{figure}[htbp]
\begin{center}
\includegraphics[width=3.0in]{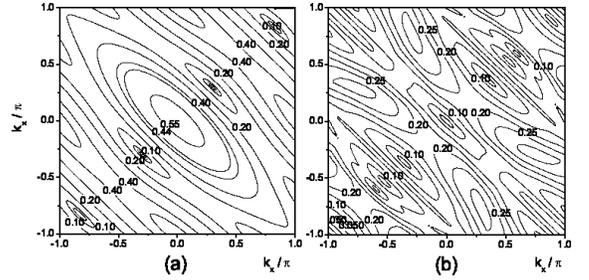}

\end{center}
\caption{(a) Contour plot of the spinon dispersion $E_{+}(k)$ as a function
of $(k_x/\pi ,k_y/\pi )$ for the $Z2A\tau ^1\tau _{+}^1\tau _{+}^3$ state.
(b) Lower edge of the two-spinon spectrum $E_{2s}(k)$ as a function of $
(k_x/\pi ,k_y/\pi )$ for the $Z2A\tau ^1\tau _{+}^1\tau _{+}^3$ state.}
\end{figure}

If the mean-field state is stable against the gauge fluctuations, we expect
that the mean-field spin-1 spectrum $E_{2s}$ should qualitatively agree with
the real spin-1 spectrum.

\section{Spin Spectral Function}
\label{sec:spe}

The quantity of interest for comparison with experiments is the spin
spectral function in $(q,\omega)$ space $S(q,\omega)$. $S(q,\omega)$ can be
calculated from the Fourier transformation of two spin correlation function
in real space $\langle \vec{S}_i(t)\cdot \vec{S}_j(0)\rangle$. Due to the
spin-rotation invariance, we obtain that 
\[
\langle \vec{S}_i\cdot \vec{S}_j\rangle =3\langle S_i^z\cdot S_j^z\rangle 
=3\langle S_i^x\cdot S_j^x\rangle =3\langle S_i^y\cdot S_j^y\rangle 
\]
The spin spectral function at zero temperature is given by 
\begin{equation}
S(q,\omega )=3S^{zz}(q,\omega )
\end{equation}
where 
\begin{equation}
S^{zz}(q,\omega ) =\sum_\lambda 2\pi |\left\langle \lambda \right|
S^z(q)\left| G\right\rangle |^2\delta (E_G+\omega -E_\lambda ), \\
\end{equation}
\begin{equation}
S^z(q) =\sum_je^{iq\cdot j}S_j^z=\frac 12( \sum_{k,\alpha }\psi _{k+q,\alpha
}^{\dagger }\psi _{k,\alpha }-N\delta_{q,0}),
\end{equation}
the summation is over all the eigenstates $\left|\lambda\right\rangle$ and $N
$ is the number of lattice sites.

Numerical calculation indicates that the neutron scattering spectrum
predicted by the dynamics spin correlation contains continuous modes besides
the main sharp peak. The Fig. 8-11 show the spin spectral function $
S^{zz}(q,\omega )$ in the q-space. The point $\Gamma $ is $(0,0)$, $M$ is $
(0,\pi )$, and $X$ is $(\pi ,\pi )$. Darker shade represents a larger
spectral function. These results can be compared with the inelastic neutron
scattering experiments. The scan from $\Gamma $ to $X$ corresponds to the
scan from $k=0$ to $k=2\pi /b$ in Fig. 2a of Ref. \cite{CTT0135}.

\begin{figure}[htbp]
\begin{center}
\includegraphics[width=3.6in]{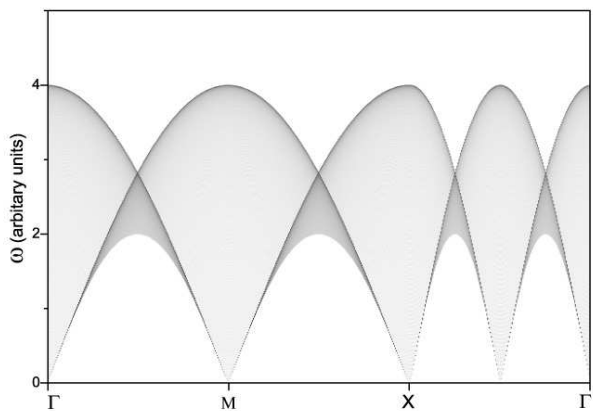}
\end{center}
\caption{ The spin spectral function $S(q,\omega)$ of the $SU^\infty(2)$
spin liquids in Eq.(\ref{Aphase}). }
\end{figure}

\begin{figure}[htbp]
\begin{center}
\includegraphics[width=3.6in]{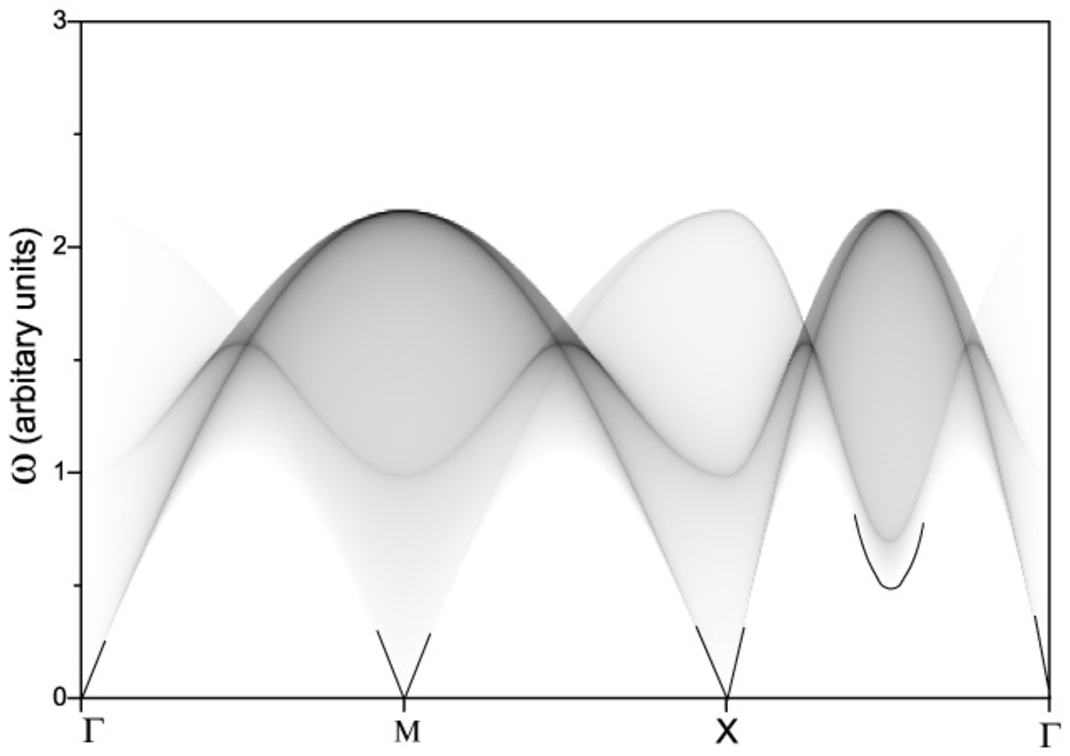}
\end{center}
\caption{ The spin spectral function $S(q,\omega)$ of the $U1B\tau ^1\tau
_{-}^0\tau _{+}^1$ state in Eq.(\ref{Bphase}). The solid lines mark the
lower edge of the spectrum.}
\end{figure}

\begin{figure}[htbp]
\begin{center}
\includegraphics[width=3.6in]{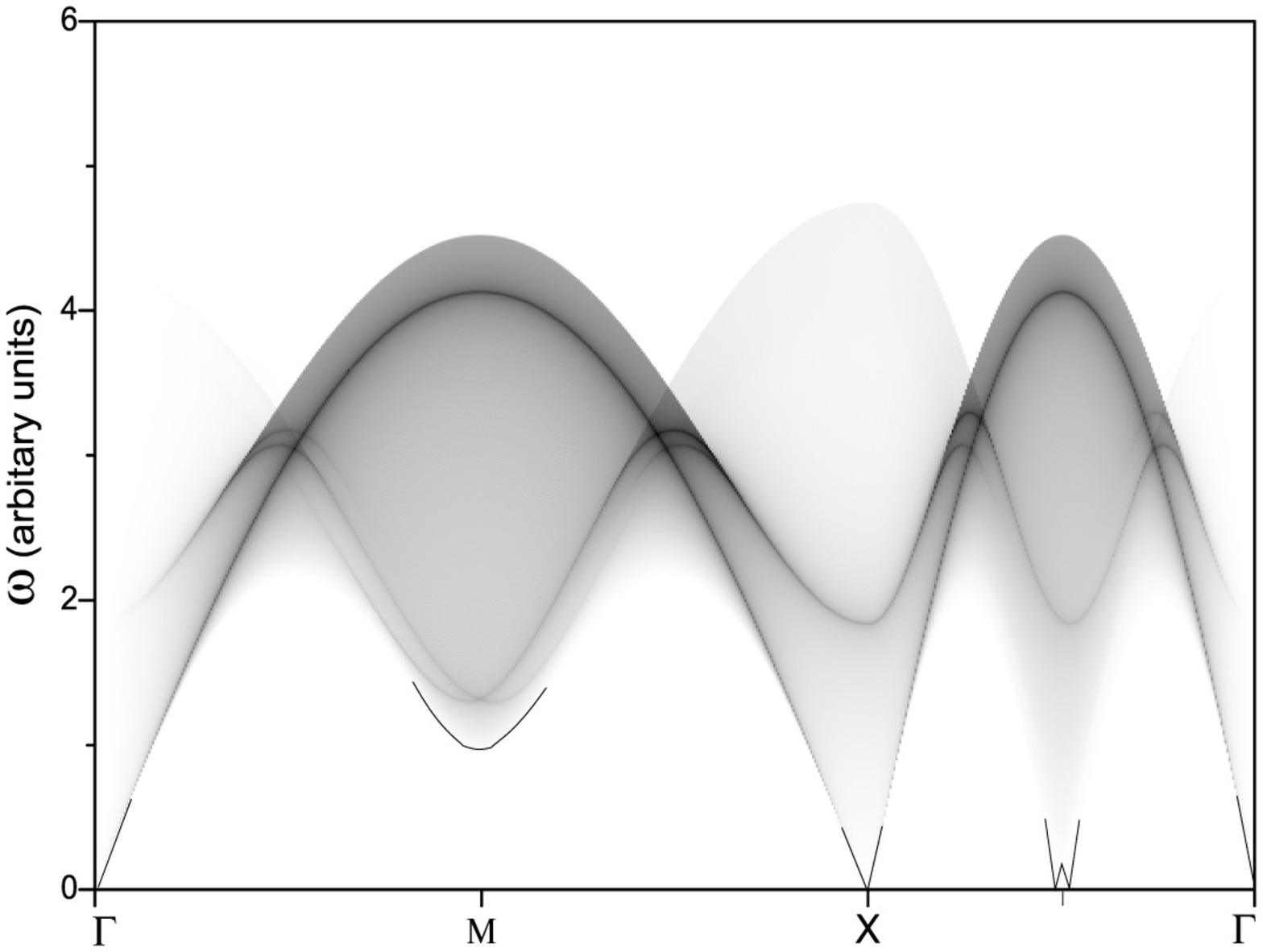}
\end{center}
\caption{The spin spectral function $S(q,\omega)$ of the $U1C\tau _{+}^0\tau
_{-}^0\tau ^1$ state in Eq.(\ref{Cphase}). The solid lines mark the lower
edge of the spectrum.}
\end{figure}

\begin{figure}[htbp]
\begin{center}
\includegraphics[width=3.6in]{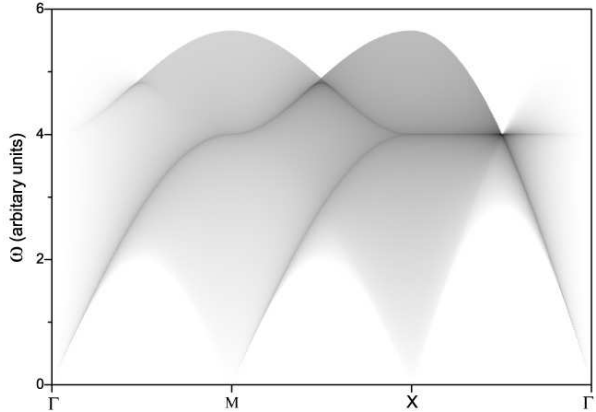}
\end{center}
\caption{ The spin spectral function $S(q,\omega)$ of the $\pi $ -flux phase
($SU2B$ state) in Eq.(\ref{Dphase})}
\end{figure}

\section{Comparison With Experiments and Previous Theoretical Results}
\label{sec:exp}

Our $J$-$J^{\prime }$ model is designed to describe the spin fluctuations in the
Cs$_2$CuCl$_4$ sample. The spin dynamics in Cs$_2$CuCl$_4$ at low temperatures
were measured by neutron scattering.\cite{CTT0135,CTH0203} To the first order
approximation, the measured spin spectral function is similar to that of
decoupled 1D spin chain (or the $SU^\infty (2)$ spin liquid) (see Fig. 8).
The decoupled 1D spin chain has gapless spin-1 excitations along the lines $
k_x+k_y=\pi $ and $k_x+k_y=2\pi $. If we examine the measured spectral
function more closely to determine the low energy boundary of spin-1
excitation $\varepsilon _1(k_x,k_y)$, we find the gapless spin-1 excitations
actually only appear at certain isolated points, implying that the spin state
is 2-dimensional. The experiment\cite{CTT0135,CTH0203} strongly suggests that there is
a gapless spin-1 excitation at $(k_x,k_y)=(\pi /2+\epsilon ,\pi /2+\epsilon )$
where $\epsilon \sim 0.04$. However, it is unclear whether the spin-1
excitation near $(k_x,k_y)=(\pi ,\pi )$ is gapless or has a small gap. The
experiment also determines the high energy boundary $\varepsilon _2(k_x,k_y)$
of spin-1 excitations. It was found that $\varepsilon _2(k_x,k_y)$ has maxima
near $(k_x,k_y)=\pm (\pi /2,\pi /2)$ (the location is not known precisely).

Using this information, we can determine which spin liquids classified in
this paper cannot describe the spin liquid state in Cs$_2$CuCl$_4$. First
the $\pi$-flux phase has no low energy spin-1 excitations near $(\pi/2,\pi/2)
$. Thus it cannot describe the spin liquid state in Cs$_2$CuCl$_4$. The $
U1B\tau ^1\tau _{-}^0\tau _{+}^1$ also does not work since the spin-1
excitations open a small gap near $(\pi/2,\pi/2)$.

It appears that the $U1C\tau _{+}^0\tau _{-}^0\tau ^1$ spin liquid and some of
its close relatives such as $Z2A\tau ^0\tau _{+}^0\tau _{+}^3$, $Z2A\tau
^1\tau _{+}^1\tau _{+}^3$, $Z2A\tau ^1\tau _{-}^3\tau _{-}^3$ and $Z2A\tau
^0\tau _{-}^1\tau _{-}^3$ spin liquids are consistent with observed spin
spectral function. Let us list the universal characters of the above five spin
liquids. The $U1C\tau _{+}^0\tau _{-}^0\tau ^1$ and $Z2A\tau ^0\tau _{+}^0\tau
_{+}^3$ spin liquids only have gapless spin-1 excitations at $(0,0) $, $(\pi
,\pi )$, $\pm (\pi /2+\epsilon ,\pi /2+\epsilon )$ and $\pm (\pi /2-\epsilon
,\pi /2-\epsilon )$. The $Z2A\tau ^1\tau _{+}^1\tau _{+}^3$ spin liquid only
has gapless spin-1 excitations at $(0,0)$, $\pm (\pi -\epsilon ^{\prime },\pi
-\epsilon ^{\prime })$, $\pm (\pi /2+\epsilon ,\pi /2+\epsilon )$ and $\pm
(\pi /2-\epsilon ,\pi /2-\epsilon )$. The $Z2A\tau ^1\tau _{-}^3\tau _{-}^3$
and $Z2A\tau ^0\tau _{-}^1\tau _{-}^3$ spin liquids only have gapless spin-1
excitations in small patches near $(0,0)$, $(\pi ,\pi )$, and $\pm (\pi
/2,\pi /2)$. By measuring the locations of gapless spin-1 excitations and
their spectral functions, we hope further experiments can determine which spin
liquid actually describes the sample.

%The $U1C\tau _{+}^0\tau _{-}^0\tau ^1$ spin liquid and its relatives also
%have strong high energy magnetic scattering near $(\pi/2,\pi /2)$ and weak
%high energy magnetic scattering near $(0,0)$ and $(\pi ,\pi )$. The
%scattering region contains multiple peaks. The upper edge of the spectrum
%has a maximum at $(\pi /2,\pi /2)$. Those properties agree very well with
%experiments.

However, to compare the spin spectral function quantitatively with
experimental measured results,\cite{CTT0135,CTH0203} we need to include
fluctuations around the mean-field state.  The interaction due to fluctuations
enhances the spin fluctuations at $(\pi/2,\pi/2)$. The strong
enough interaction can drive the transition from the spin liquid state to the
spiral state.

\begin{figure}[htbp]
\begin{center}
\includegraphics[width=3.1in]{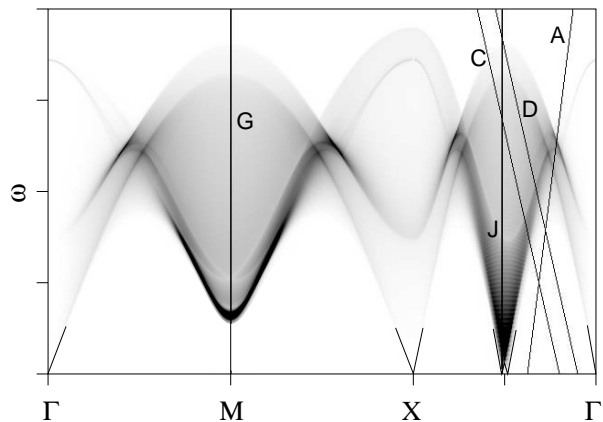}
\end{center}
\caption{The spin spectral function $S(q,\omega)$ of the $U1C\tau _{+}^0\tau
_{-}^0\tau ^1$ state obtained by RPA. The short solid lines mark the lower
edge of the spectrum.
\label{imgRPA}
}
\end{figure}

\begin{figure}[htbp]
\begin{center}
\includegraphics[width=3.1in]{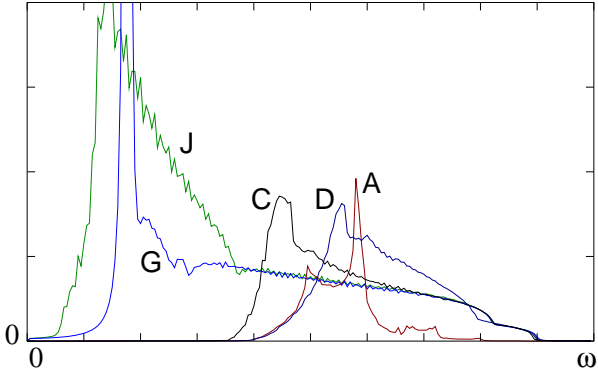}
\end{center}
\caption{The RPA spin spectral function $S(q,\omega)$ along the lines
A, C, D, G, and J in Fig. \ref{imgRPA} for the $U1C\tau _{+}^0\tau
_{-}^0\tau ^1$ state.
\label{imgACDGJ}
}
\end{figure}

To include the interaction enhancement of the $(\pi/2,\pi/2)$ spin fluctuations,
we consider the effects of the $J$-term (the diagonal coupling term):
$\sum_{\mathbf{i}} J\mathbf{S}_{\mathbf{i}} \cdot \mathbf{S}_{\mathbf{i}
+\mathbf{x}+\mathbf{y}}$.
Within the random phase approximation (RPA), the spin correlation function
$\pi(\mathbf{k}, \mathbf{\omega})$ can be obtained from the
mean-field spin correlation function
$\pi_0(\mathbf{k}, \mathbf{\omega})$:
\begin{equation}
 \pi(\mathbf{k}, \mathbf{\omega})
=\frac{\pi_0(\mathbf{k}, \mathbf{\omega})}
      {1-\tilde J\cos(k_x+k_y)\pi_0(\mathbf{k}, \mathbf{\omega})}
\end{equation}
The imaginary part of $\pi(\mathbf{k}, \mathbf{\omega})$ gives us the RPA spin
spectral function.  To be definite, we start with the mean-field $U1C\tau
_{+}^0\tau _{-}^0\tau ^1$ spin liquid state and choose $\pi_0(\mathbf{k},
\mathbf{\omega})$ to be the mean-field spin correlation function of the
$U1C\tau _{+}^0\tau _{-}^0\tau ^1$ state.  The
experiments\cite{CTT0135,CTH0203}
indicate that the system is close to the spiral phase. So here we choose the
value of $\tilde J$ to make the system close to the spiral phase instability.
The resulting RPA spin spectral function is plotted in Fig. \ref{imgRPA} and
Fig.  \ref{imgACDGJ}.  The spin spectral functions along the lines A, C, and D
in Fig. \ref{imgRPA} were measured by experiments.\cite{CTT0135} The simple
RPA results are quite similar to the measured results (see Fig. 3 in Ref.
\cite{CTT0135}).

We would like to remark that the low energy effective theory for the $U1C\tau
_{+}^0\tau _{-}^0\tau ^1$ state contain gapless fermions interacting with
$U(1)$ gauge field. Depending on that strength of the $U(1)$ gauge
fluctuations, the $U(1)$ gauge interaction may or may not destabilize the
$U1C\tau _{+}^0\tau _{-}^0\tau ^1$ state.  It is shown that if the spinons has
a very anisotropic dispersion, the instanton effect of the 2+1 $U(1)$ gauge
theory is irrelevant.\cite{Wqoslpub} In this case, the $U1C\tau _{+}^0\tau
_{-}^0\tau ^1$ state is an algebraic spin liquid,\cite{RWspin} which contains
gapless spin-1 excitations, but non of the gapless excitations are described
by well defined quasiparticles.  If the spinons has a more isotropic
dispersion, the instanton effect is relevant, which lead to a confinement of
the $U(1)$ gauge field. In this case, the $U1C\tau _{+}^0\tau _{-}^0\tau ^1$
state may become one of the stable $Z_2$ spin liquid states or other symmetry
breaking states (such as the spiral state) below a transition temperature $T_c$.

Even when the $U1C\tau _{+}^0\tau _{-}^0\tau ^1$ state is unstable, it is
still useful for understanding the system at finite temperatures.  The $U1C\tau
_{+}^0\tau _{-}^0\tau ^1$ state may control the dynamical properties of the
Cs$_2$CuCl$_4$ samples above the $T_c$, provided that $T_c$ is much smaller
than the band width of the spin fluctuations.

\section{Conclusion}
\label{sec:con}

In this paper, we use the projective symmetry group (PSG) to characterize
quantum orders in symmetric spin liquids on 2D triangular lattice. We classify
these symmetric spin liquids through PSG, and find that there are 63 kinds of
$Z_2$ spin liquids, 30 kinds of $U(1)$ spin liquids and 2 kinds of $SU(2)$
spin liquids. The mean-field phase diagram Fig. 2. for the nearest neighbor
spin coupling system is calculated. One-spinon and two-spinon excitation
spectrum for some $Z_2$ and $U(1)$ spin liquids are inspected. We show that
the gapless spinon-1 excitation spectrum can be used to physically measure the
quantum order. By examining the measured spin spectral function in
Cs$_2$CuCl$_4$ sample,\cite{CTT0135} we find that it is possible that the spin
liquid state in Cs$_2$CuCl$_4$ is described by the $U1C\tau _{+}^0\tau
_{-}^0\tau ^1$ spin liquid or one of its relatives. 
%We discussed the characters of those spin liquids so that future experiments
%can determine which spin liquids actually describe the spin state in
%Cs$_2$CuCl$_4$.

We would like to point out that the RPA spin spectral function for the
$U1C\tau _{+}^0\tau _{-}^0\tau ^1$ spin liquid is very close to the spin
spectral function\cite{CVK0352} obtained from the slave-fermion
approach.\cite{S9277,CMM0159} In the experiments,\cite{CTT0135,CTH0203} the
spin liquid was observed only at a low but finite temperature. At that
temperature, it is hard to distinguish to the two proposed states.  However, if
we find a spin liquid state at zero temperature, then we can tell which of the
two proposed spin liquids can describe the experiments.  This is because at
zero temperature, the  $U1C\tau _{+}^0\tau _{-}^0\tau ^1$ spin liquid (and its
relatives) and the spin liquid from the slave-fermion approach are
qualitatively different.  The $U1C\tau _{+}^0\tau _{-}^0\tau ^1$ spin liquid
(and its relatives) contains gapless spin-1 excitations while the spin liquid
from the slave-fermion approach is fully gapped.

XGW is supported in part by NSF Grant No. DMR--01--23156
and by NSF-MRSEC Grant No. DMR--02--13282.

\appendix

\section{Classification of Z$_2$ Spin Liquids}

In this appendix we consider the connected ansatz only, which is interesting
and cover a wide range of spin liquids in fact. For any two sites $i$ and 
$j$, a connected ansatz will offer a path $(i_1i_2\cdots i_n)$, which satisfy
that $u_{i_k,i_{k+1}}\neq 0$, $k=1,2,...,n$, $i_1=i$ and $i_n=j$. Due to the
translation symmetry of the ansatz, we can choose a gauge in which all the
loop operators of the ansatz are translation invariant. We will call such a
gauge uniform gauge. Hereafter we will work in uniform gauge. According to
the definition of PSG and uniform gauge, 
\begin{eqnarray*}
G_x(i)P_{T_xC}G_x^{\dagger }(i) &=&G_x(i)P_CG_x^{\dagger }(i)=P_C, \\
G_x(i)P_C &=&P_CG_x(i).
\end{eqnarray*}
For $Z_2$ spin liquids, different loop operators basing at the same base
point do not commute. We see that translation invariance of $P_C$ in the
uniform gauge requires that 
\[
G_x(i),G_y(i)\in \mathcal{G}=\left\{ \pm \tau ^0\right\} . 
\]
Gauge transformations $W(i)=\left( -\right) ^{f(i)}\tau ^0$ $(f(i)=\pm 1)$
do not change the translation invariant property of the loop operators. We
can always choose a gauge which satisfy that : 
\begin{equation}
G_y(i)=\tau ^0.  \label{Gy}
\end{equation}
Gauge transformation $W(i)=W(i_x)$ does not change the condition $
G_y(i)=\tau ^0$, hence we can gauge fix that 
\begin{equation}
G_x\left( i_x,i_y=0\right) =\tau ^0.  \label{Gx}
\end{equation}
From the relations (\ref{TxTy}), it is easy to see that 
\begin{equation}
G_x\left( i\right) G_y\left( i-\hat{x}\right) G_x^{-1}\left( i-\hat{y}
\right) G_y^{-1}\left( i\right) =\pm \tau ^0.
\end{equation}
Substituting Eqs. (\ref{Gy}) and (\ref{Gx}) into the above, one sees that 
\begin{equation}
G_x\left( i\right) G_x^{-1}\left( i-\hat{y}\right) =\pm \tau ^0.
\end{equation}
Therefore there are two classes of PSG belong to translation symmetry, 
\begin{equation}
G_x\left( i\right) =G_y\left( i\right) =\tau ^0
\end{equation}
and 
\begin{equation}
G_x\left( i\right) =\left( -\right) ^{i_y}\tau ^0,G_y\left( i\right) =\tau
^0.
\end{equation}
We will label the former $Z2A$ and the later $Z2B$.

Then we consider the spin parity symmetry $T^{*}$ which satisfies that 
\begin{equation}
T^{*}T_xT^{*-1}T_x^{-1}=T^{*}T_yT^{*-1}T_y^{-1}=T^{*2}=\pm 1.
\end{equation}
In a similar way, we have the following relations 
\begin{eqnarray}
G_{T^{*}}T^{*}G_xT_x(G_{T^{*}}T^{*})^{-1}(G_xT_x)^{-1} &\in &\mathcal{G}, 
\nonumber \\
G_{T^{*}}T^{*}G_yT_y(G_{T^{*}}T^{*})^{-1}(G_yT_y)^{-1} &\in &\mathcal{G},
\label{T*Txy} \\
G_{T^{*}}T^{*}G_{T^{*}}T^{*} &\in &\mathcal{G},  \nonumber
\end{eqnarray}
then 
\begin{eqnarray}
G_{T^{*}}\left( i\right) G_x\left( i\right) G_{T^{*}}^{-1}\left( i-\hat{x}
\right) G_x^{-1}\left( i\right) &=&\eta _{xt}\tau ^0,  \nonumber \\
G_{T^{*}}\left( i\right) G_y\left( i\right) G_{T^{*}}^{-1}\left( i-\hat{y}
\right) G_y^{-1}\left( i\right) &=&\eta _{yt}\tau ^0, \\
G_{T^{*}}\left( i\right) ^2 &=&\pm \tau ^0,  \nonumber
\end{eqnarray}
\begin{equation}
G_{T^{*}}\left( i\right) =\eta _{xt}^{i_x}\eta _{yt}^{i_y}g_{T^{*}},
\end{equation}
where $\eta _{xt}=\pm 1$, $\eta _{yt}=\pm 1$ and $g_{T^{*}}^2=\pm \tau ^0$.
Two gauge inequivalent choices of $g_{T^{*}}$ are $g_{T^{*}}=\tau ^0$ and $
g_{T^{*}}=i\tau ^3$.

Next we add two types parity transformations $P_{xy}\left( i_x,i_y\right)
=\left( i_y,i_x\right) $ and $P_{x\bar{y}}\left( i_x,i_y\right) =\left(
-i_y,-i_x\right) $, which satisfy that 
\begin{eqnarray}
T_xP_{xy}T_y^{-1}P_{xy}^{-1} &=&1,  \nonumber \\
T_yP_{xy}T_x^{-1}P_{xy}^{-1} &=&1,  \nonumber \\
T_xP_{x\bar{y}}T_yP_{x\bar{y}} &=&1,  \nonumber \\
T_xP_{x\bar{y}}T_yP_{x\bar{y}} &=&1, \\
T^{*}P_{xy}T^{*-1}P_{xy}^{-1} &=&1,  \nonumber \\
T^{*}P_{x\bar{y}}T^{*-1}P_{x\bar{y}} &=&1,  \nonumber
\end{eqnarray}
From the above equations, we find that 
\begin{eqnarray}  \label{PxyT*Txy}
(G_xT_x)(G_{P_{xy}}P_{xy})(G_yT_y)^{-1}(G_{P_{xy}}P_{xy})^{-1} &\in &
\mathcal{G}  \nonumber \\
(G_yT_y)(G_{P_{xy}}P_{xy})(G_xT_x)^{-1}(G_{P_{xy}}P_{xy})^{-1} &\in &
\mathcal{G}  \nonumber \\
(G_xT_x)(G_{P_{x\bar{y}}}P_{x\bar{y}})(G_yT_y)(G_{P_{x\bar{y}}}P_{x\bar{y}
})^{-1} &\in &\mathcal{G}  \nonumber \\
(G_yT_y)(G_{P_{x\bar{y}}}P_{x\bar{y}})(G_xT_x)(G_{P_{x\bar{y}}}P_{x\bar{y}
})^{-1} &\in &\mathcal{G}  \nonumber \\
G_{T^{*}}T^{*}(G_{P_{xy}}P_{xy})(G_{T^{*}}T^{*})^{-1}(G_{P_{xy}}P_{xy})^{-1}
&\in &\mathcal{G}  \nonumber \\
G_{T^{*}}T^{*}(G_{P_{x\bar{y}}}P_{x\bar{y}})(G_{T^{*}}T^{*})^{-1}(G_{P_{x
\bar{y}}}P_{x\bar{y}})^{-1} &\in &\mathcal{G}  \nonumber \\
\end{eqnarray}
and 
\begin{eqnarray}
G_x\left( i\right) G_{P_{xy}}\left( i-\hat{x}\right) G_y^{-1}\left(
P_{xy}\left( i\right) \right) G_{P_{xy}}^{-1}\left( i\right) &=&\eta
_{xpxy}\tau ^0  \nonumber \\
G_y\left( i\right) G_{P_{xy}}\left( i-\hat{y}\right) G_x^{-1}\left(
P_{xy}\left( i\right) \right) G_{P_{xy}}^{-1}\left( i\right) &=&\eta
_{ypxy}\tau ^0  \nonumber \\
G_x\left( i\right) G_{P_{x\bar{y}}}\left( i-\hat{x}\right) G_y^{-1}\left(
P_{x\bar{y}}\left( i\right) -\hat{x}\right) G_{P_{x\bar{y}}}^{-1}\left(
i\right) &=&\eta _{xpx\bar{y}}\tau ^0  \nonumber \\
G_y\left( i\right) G_{P_{x\bar{y}}}\left( i-\hat{y}\right) G_x^{-1}\left(
P_{x\bar{y}}\left( i\right) -\hat{y}\right) G_{P_{x\bar{y}}}^{-1}\left(
i\right) &=&\eta _{ypx\bar{y}}\tau ^0  \nonumber \\
G_{T^{*}}\left( i\right) G_{P_{xy}}\left( i\right) G_{T^{*}}^{-1}\left(
P_{xy}\left( i\right) \right) G_{P_{xy}}^{-1}\left( i\right) &=&\eta
_{tpxy}\tau ^0  \nonumber \\
G_{T^{*}}\left( i\right) G_{P_{x\bar{y}}}\left( i\right)
G_{T^{*}}^{-1}\left( P_{x\bar{y}}\left( i\right) \right) G_{P_{x\bar{y}
}}^{-1}\left( i\right) &=&\eta _{tpx\bar{y}}\tau ^0  \nonumber \\
&&
\end{eqnarray}
where $\eta _{xpxy}=\pm 1$...

For $Z2A$ type PSG's, 
\begin{eqnarray}
G_{P_{xy}}\left( i\right) &=&\eta _{xpxy}^{i_x}\eta _{ypxy}^{i_y}g_{P_{xy}},
\nonumber \\
G_{P_{x\bar{y}}}\left( i\right) &=&\eta _{xpx\bar{y}}^{i_x}\eta _{ypx\bar{y}
}^{i_y}g_{P_{x\bar{y}}}.
\end{eqnarray}

For $Z2B$ type PSG's, 
\begin{eqnarray}
G_{P_{xy}}\left( i\right) &=&\left( -\right) ^{i_xi_y}\eta _{xpxy}^{i_x}\eta
_{ypxy}^{i_y}g_{P_{xy}},  \nonumber \\
G_{P_{x\bar{y}}}\left( i\right) &=&\left( -\right) ^{i_xi_y}\eta _{xpx\bar{y}
}^{i_x}\eta _{ypx\bar{y}}^{i_y}g_{P_{x\bar{y}}}.
\end{eqnarray}

We also have that 
\begin{equation}
P_{xy}^2=P_{x\bar{y}}^2=P_{xy}P_{x\bar{y}}P_{xy}^{-1}P_{x\bar{y}}^{-1}=1
\label{Pxy}
\end{equation}
which leads to 
\begin{eqnarray*}
G_{P_{xy}}\left( i\right) G_{P_{xy}}\left( P_{xy}\left( i\right) \right)
&=&\pm \tau ^0, \\
G_{P_{x\bar{y}}}\left( i\right) G_{P_{x\bar{y}}}\left( P_{x\bar{y}}\left(
i\right) \right) &=&\pm \tau ^0,
\end{eqnarray*}
and 
\begin{eqnarray*}
\left( \eta _{xpxy}\eta _{ypxy}\right) ^{i_x+i_y}g_{P_{xy}}^2 &=&\pm \tau ^0,
\\
\left( \eta _{xpx\bar{y}}\eta _{ypx\bar{y}}\right) ^{i_x+i_y}g_{P_{x\bar{y}
}}^2 &=&\pm \tau ^0,
\end{eqnarray*}
The above implies that 
\begin{eqnarray*}
\eta _{xpxy} &=&\eta _{ypxy}=\eta _{pxy}, \\
\eta _{xpx\bar{y}} &=&\eta _{ypx\bar{y}}=\eta _{px\bar{y}}, \\
g_{P_{xy}}^2 &=&\pm \tau ^0,g_{P_{x\bar{y}}}^2=\pm \tau ^0.
\end{eqnarray*}
By the similar way, one sees that 
\begin{equation}
\eta _{xt}=\eta _{yt}=\eta _t.
\end{equation}
Gauge transformation $W(i)=\left( -\right) ^{i_x}$ will change $\eta _{pxy}$
to $-\eta _{pxy}$, therefore we can chose a gauge in which $\eta _{pxy}=1$.

In conclusion, we have the following 60 kinds of $Z2A$ algebraic PSG's and
60 kinds of $Z2B$ algebraic PSG's on our triangular lattices:

$Z2Ag_{P_{xy}}(g_{P_{x\bar{y}}})_{\eta _{px\bar{y}}}(g_{T^{*}})_{\eta _t}$
\begin{eqnarray}
G_x(i) &=&G_y(i)=\tau ^0,  \nonumber \\
G_{T^{*}}(i) &=&\eta _t^ig_{T^{*}},\eta _t=\pm 1,  \nonumber \\
G_{P_{xy}}(i) &=&g_{P_{xy}}, \\
G_{P_{x\bar{y}}}(i) &=&\eta _{px\bar{y}}^ig_{P_{x\bar{y}}},\eta _{px\bar{y}
}=\pm 1,  \nonumber
\end{eqnarray}
and

$Z2Bg_{P_{xy}}(g_{P_{x\bar{y}}})_{\eta _{px\bar{y}}}(g_{T^{*}})_{\eta _t}$
\begin{eqnarray}
G_y(i) &=&\tau ^0,G_x(i)=\left( -\right) ^{i_y}\tau ^0,  \nonumber \\
G_{T^{*}}(i) &=&\eta _t^ig_{T^{*}},\eta _t=\pm 1,  \nonumber \\
G_{P_{xy}}(i) &=&\left( -\right) ^{i_xi_y}g_{P_{xy}}, \\
G_{P_{x\bar{y}}}(i) &=&\left( -\right) ^{i_xi_y}\eta _{px\bar{y}}^ig_{P_{x
\bar{y}}},\eta _{px\bar{y}}=\pm 1,  \nonumber
\end{eqnarray}
where $g_{T^{*}}$, $g_{P_{xy}}$, and $g_{P_{x\bar{y}}}$ satisfy that 
\begin{eqnarray}
g_{T^{*}}^2 &=&\pm \tau ^0,  \nonumber \\
g_{P_{xy}}^2 &=&\pm \tau ^0,  \nonumber \\
g_{P_{x\bar{y}}}^2 &=&\pm \tau ^0,  \nonumber \\
g_{T^{*}}g_{P_{xy}}g_{T^{*}}^{-1}g_{P_{xy}}^{-1} &=&\pm \tau ^0, \\
g_{T^{*}}g_{P_{x\bar{y}}}g_{T^{*}}^{-1}g_{P_{x\bar{y}}}^{-1} &=&\pm \tau ^0,
\nonumber \\
g_{P_{xy}}g_{P_{x\bar{y}}}g_{P_{xy}}^{-1}g_{P_{x\bar{y}}}^{-1} &=&\pm \tau
^0.  \nonumber
\end{eqnarray}

All the gauge inequivalent $g_{T^{*}}$, $g_{P_{xy}}$, and $g_{P_{x\bar{y}}}$
are given by the following table

\[
\begin{tabular}{|c|c|c|c|}
\hline
$g_{T^{*}}$ & $g_{P_{xy}}$ & $g_{P_{x\bar{y}}}$ & $g_{P_{xy}}g_{P_{x\bar{y}
}} $ \\ \hline
$\tau ^0$ & $\tau ^0$ & $\tau ^0$ & $\tau ^0$ \\ \hline
$\tau ^0$ & $\tau ^0$ & $i\tau ^3$ & $i\tau ^3$ \\ \hline
$\tau ^0$ & $i\tau ^3$ & $\tau ^0$ & $i\tau ^3$ \\ \hline
$\tau ^0$ & $i\tau ^3$ & $i\tau ^3$ & $\tau ^0$ \\ \hline
$\tau ^0$ & $i\tau ^3$ & $i\tau ^1$ & $i\tau ^2$ \\ \hline
$i\tau ^3$ & $\tau ^0$ & $\tau ^0$ & $\tau ^0$ \\ \hline
$i\tau ^3$ & $\tau ^0$ & $i\tau ^3$ & $i\tau ^3$ \\ \hline
$i\tau ^3$ & $\tau ^0$ & $i\tau ^1$ & $i\tau ^1$ \\ \hline
$i\tau ^3$ & $i\tau ^3$ & $\tau ^0$ & $i\tau ^3$ \\ \hline
$i\tau ^3$ & $i\tau ^3$ & $i\tau ^3$ & $\tau ^0$ \\ \hline
$i\tau ^3$ & $i\tau ^3$ & $i\tau ^1$ & $i\tau ^2$ \\ \hline
$i\tau ^3$ & $i\tau ^1$ & $\tau ^0$ & $i\tau ^1$ \\ \hline
$i\tau ^3$ & $i\tau ^1$ & $i\tau ^3$ & $\tau ^2$ \\ \hline
$i\tau ^3$ & $i\tau ^1$ & $i\tau ^1$ & $\tau ^0$ \\ \hline
$i\tau ^3$ & $i\tau ^1$ & $i\tau ^2$ & $i\tau ^3$ \\ \hline
\end{tabular}
\]

After finding all the PSGs, next we need to find the ansatz that is
invariant under those PSGs. This way we classify (mean-field) $Z_2$ spin
liquids through PSG's. Let us first consider the translation symmetry. We
have two classes of spin liquids,

$Z2A$ spin liquids 
\begin{eqnarray}
G_y(i) &=&\tau ^0,G_x(i)=\tau ^0,  \nonumber \\
u_{i,i+m} &=&u_m,
\end{eqnarray}
and

$Z2B$ spin liquids 
\begin{eqnarray}
G_y(i) &=&\tau ^0,G_x(i)=\left( -\right) ^{i_y}\tau ^0,  \nonumber \\
u_{i,i+m} &=&\left( -\right) ^{m_yi_x}u_m.
\end{eqnarray}

Considering the spin parity symmetry $T^{*}$, we have

\begin{equation}
G_{T^{*}}T^{*}\left( u_{i,i+m}\right) =u_{i,i+m},  \label{TSpin}
\end{equation}
which reads

\begin{equation}
-\eta _t^mg_{T^{*}}u_mg_{T^{*}}^{-1}=u_m
\end{equation}

Therefore, for different $(g_{T^{*}})_{\eta _t}$, we have the following
table:

\[
\begin{tabular}{|c|c|}
\hline
$\tau _{+}^0$ & $u_m=0$ \\ \hline
$\tau _{-}^0$ & $u_m=0,$if $m=even$ \\ \hline
$\tau _{+}^3$ & $u_m=u_m^1\tau ^1+u_m^2\tau ^2$ \\ \hline
$\tau _{-}^3$ & 
\begin{tabular}{cc}
$u_m=u_m^1\tau ^1+u_m^2\tau ^2,$ & if $m=even$ \\ 
$u_m=u_m^0\tau ^0+u_m^3\tau ^3,$ & if $m=odd$
\end{tabular}
\\ \hline
\end{tabular}
\]

Now we consider the 180$^{\circ }$ rotation symmetry, which can be
constructed through combining $P_{xy}$ and $P_{x\bar{y}}$,

\begin{equation}
G_{P_{xy}}P_{xy}G_{P_{x\bar{y}}}P_{x\bar{y}}(u_{i,i+m})=u_{i,i+m}
\label{180rotation}
\end{equation}
the above leads to

\begin{equation}
\eta _{_{px\bar{y}}}^mg_{P_{xy}}g_{P_{x\bar{y}}}u_mg_{P_{x\bar{y}
}}^{-1}g_{P_{xy}}^{-1}=u_m^{\dagger },
\end{equation}
for the $Z2A$ spin liquids, and

\begin{equation}
\left( -\right) ^{m_xm_y}\eta _{_{px\bar{y}}}^mg_{P_{xy}}g_{P_{x\bar{y}
}}u_mg_{P_{x\bar{y}}}^{-1}g_{P_{xy}}^{-1}=u_m^{\dagger }.
\end{equation}
for the $Z2B$ spin liquids.

By writing all the ansatz, we find that although there are 120 kinds $Z_2$
algebraic PSG's, only 23 of them lead to $Z2A$ spin liquids and 40 of them
lead to $Z2B$ spin liquids. Other algebraic PSG's lead to vanishing ansatz, $
U(1)$ or $SU(2)$ spin liquids.

\section{Classification of $U(1)$ Spin Liquids}

For $U(1)$ spin liquids, we can choose a gauge so that $u_{ij}$ take the
form 
\begin{equation}
u_{ij}=i\rho _{ij}e^{i\theta _{ij}\tau ^3}=u_{ij}^0\tau ^0+u_{ij}^3\tau ^3
\end{equation}
We will call this gauge canonical gauge. In the canonical gauge, IGG has a
form of $\mathcal{G}=\left\{ e^{i\theta \tau ^3},\theta \in [0,2\pi
)\right\} ,$ where $\theta _i$ is real for each $i$. Due to the translation
symmetry of the ansatz, the loop operator $P_C$ have a form 
\begin{equation}
P_{T_xC}=\left( \tau ^1\right) ^{n_i}P_C\left( \tau ^1\right) ^{n_i}
\end{equation}
where $n_i=0,1$. The gauge transformation $G_{x,y}$ associated with the
translation take the following form 
\begin{eqnarray}
G_x(i) &=&\left( -i\tau ^1\right) ^{n_i}e^{i\theta _x(i)\tau ^3}\left( i\tau
^1\right) ^{n_{i-\hat{x}}}  \nonumber \\
G_y(i) &=&\left( -i\tau ^1\right) ^{n_i}e^{i\theta _y(i)\tau ^3}\left( i\tau
^1\right) ^{n_{i-\hat{y}}}
\end{eqnarray}
in the canonical gauge. We note that a gauge transformations keep $u_{ij}$
to have the form in the canonical gauge must have one of the following two
forms 
\begin{eqnarray}
W_i &=&e^{i\theta _i\tau ^3}  \label{U1Gauge} \\
W_i &=&e^{i\theta _i\tau ^3}\left( i\tau ^1\right)  \label{U1Gauge2}
\end{eqnarray}
For spin liquids with connected $u_{ij}$, $G_{x,y}$ must take one of the
above two forms in the canonical gauge. Thus $n_i$ can only be one of the
following four choices: $n_i=0,n_i=(1-(-)^i)/2,n_i=(1-(-)^{i_x})/2$, and $
n_i=(1-(-)^{i_y})/2$. In these four cases, $G_{x,y}$ take one of the above
two forms and $u_{ij}$ can be connected.

Let us consider those cases in turn. We will work in the canonical gauge.

When $n_i=0$, $G_{x,y}$ have a form 
\begin{eqnarray}
G_x(i) &=&e^{i\theta _x(i)\tau ^3}  \nonumber \\
G_y(i) &=&e^{i\theta _y(i)\tau ^3}
\end{eqnarray}
by (\ref{U1Gauge}) we can gauge fix them up to $\theta _x(i_x,0)=0$ and $
\theta _y(i)=0$. The relation (\ref{TxTy}) requires that 
\begin{equation}
\theta _x\left( i\right) +\theta _y\left( i-\hat{x}\right) -\theta _x\left(
i-\hat{y}\right) -\theta _y\left( i\right) =\varphi
\end{equation}
Therefore $G_{x,y}$ have a form 
\begin{eqnarray}
G_x(i) &=&e^{i(i_y\varphi +\theta _x)\tau ^3}  \nonumber \\
G_y(i) &=&e^{i\theta _y\tau ^3}
\end{eqnarray}
The ansatz with translation symmetry has a form 
\begin{equation}
u_{i,i+m}=i\rho _me^{i(-m_yi_x\varphi +\phi _m)\tau ^3}  \label{U1ABansatz}
\end{equation}

When $n_i=(1-(-)^i)/2$, 
\begin{eqnarray}
G_x(i) &=&e^{i\theta _x(i)\tau ^3}(i\tau ^1)  \nonumber \\
G_y(i) &=&e^{i\theta _y(i)\tau ^3}(i\tau ^1)
\end{eqnarray}
then 
\begin{eqnarray}
G_x(i) &=&e^{i((-)^{i_y}\varphi +\theta _x)\tau ^3}(i\tau ^1)  \nonumber \\
G_y(i) &=&e^{i\theta _y\tau ^3}(i\tau ^1)
\end{eqnarray}
Using gauge transformation $W_i=e^{i((-)^{i_y-1}\varphi /2)\tau ^3}$, the
above change to 
\begin{eqnarray}
G_x(i) &=&e^{i\theta _x\tau ^3}(i\tau ^1)  \nonumber \\
G_y(i) &=&e^{i\theta _y\tau ^3}(i\tau ^1)
\end{eqnarray}
The ansatz with translation symmetry has a form 
\begin{equation}
u_{i,i+m}=i\rho _me^{i(-)^i\phi _m\tau ^3}
\end{equation}

When $n_i=(1-(-)^{i_x})/2$, 
\begin{eqnarray}
G_x(i) &=&e^{i\theta _x(i)\tau ^3}(i\tau ^1)  \nonumber \\
G_y(i) &=&e^{i\theta _y(i)\tau ^3}
\end{eqnarray}
then 
\begin{eqnarray}
G_x(i) &=&e^{i(i_y\varphi +\theta _x)\tau ^3}(i\tau ^1)  \nonumber \\
G_y(i) &=&e^{i\theta _y\tau ^3}
\end{eqnarray}
Using gauge transformation $W_i=e^{i(-i_y\varphi /2)\tau ^3}$, the above
change to 
\begin{eqnarray}
G_x(i) &=&e^{i\theta _x\tau ^3}(i\tau ^1)  \nonumber \\
G_y(i) &=&e^{i\theta _y\tau ^3}
\end{eqnarray}
The ansatz with translation symmetry has a form 
\begin{equation}
u_{i,i+m}=i\rho _me^{i(-)^{i_x}\phi _m\tau ^3}
\end{equation}

When $n_i=(1-(-)^{i_y})/2$, 
\begin{eqnarray}
G_x(i) &=&e^{i\theta _x(i)\tau ^3}  \nonumber \\
G_y(i) &=&e^{i\theta _y(i)\tau ^3}(i\tau ^1)
\end{eqnarray}
then 
\begin{eqnarray}
G_x(i) &=&e^{i((-)^{i_y}\varphi +\theta _x)\tau ^3}  \nonumber \\
G_y(i) &=&e^{i\theta _y\tau ^3}(i\tau ^1)
\end{eqnarray}
Using gauge transformation $W_i=e^{i((-)^{i_y-1}\varphi /2)\tau ^3}$, the
above change to 
\begin{eqnarray}
G_x(i) &=&e^{i\theta _x\tau ^3}  \nonumber \\
G_y(i) &=&e^{i\theta _y\tau ^3}(i\tau ^1)
\end{eqnarray}
The ansatz with translation symmetry has a form 
\begin{equation}
u_{i,i+m}=i\rho _me^{i(-)^{i_y}\phi _m\tau ^3}
\end{equation}

Now we add parity symmetry $P_{xy}$. Considering the flux in a plaquette,
the parity symmetry requires $\varphi =0$ or $\pi $ in (\ref{U1ABansatz}).
From the relations 
\begin{eqnarray}
G_x\left( i\right) G_{P_{xy}}\left( i-\hat{x}\right) G_y^{-1}\left(
P_{xy}\left( i\right) \right) G_{P_{xy}}^{-1}\left( i\right) &=&e^{i\theta
_{xpxy}\tau ^3}  \nonumber \\
G_y\left( i\right) G_{P_{xy}}\left( i-\hat{y}\right) G_x^{-1}\left(
P_{xy}\left( i\right) \right) G_{P_{xy}}^{-1}\left( i\right) &=&e^{i\theta
_{ypxy}\tau ^3}  \nonumber \\
G_x\left( i\right) G_{P_{x\bar{y}}}\left( i-\hat{x}\right) G_y^{-1}\left(
P_{x\bar{y}}\left( i\right) -\hat{x}\right) G_{P_{x\bar{y}}}^{-1}\left(
i\right) &=&e^{i\theta _{xpx\bar{y}}\tau ^3}  \nonumber \\
G_y\left( i\right) G_{P_{x\bar{y}}}\left( i-\hat{y}\right) G_x^{-1}\left(
P_{x\bar{y}}\left( i\right) -\hat{y}\right) G_{P_{x\bar{y}}}^{-1}\left(
i\right) &=&e^{i\theta _{ypx\bar{y}}\tau ^3}\   \nonumber \\
G_{P_{xy}}\left( i\right) G_{P_{xy}}\left( P_{xy}\left( i\right) \right)
&=&e^{i\theta _{pxy}\tau ^3}  \nonumber \\
G_{P_{x\bar{y}}}\left( i\right) G_{P_{x\bar{y}}}\left( P_{x\bar{y}}\left(
i\right) \right) &=&e^{i\theta _{px\bar{y}}\tau ^3}  \nonumber \\
G_{P_{xy}}\left( i\right) G_{P_{x\bar{y}}}\left( P_{xy}\left( i\right)
\right) G_{P_{xy}}^{-1}\left( P_{x\bar{y}}\left( i\right) \right) G_{P_{x
\bar{y}}}^{-1}\left( i\right) &=&e^{i\theta _p\tau ^3}  \nonumber \\
&&  \label{PxyPxy}
\end{eqnarray}
we will find only $n_i=0$ and $n_i=(1-(-)^i)/2$ lead to the existence of $
G_{P_{xy}}\left( i\right) $. Then we can classify the $U(1)$ spin liquids
with parity symmetry into three classes according to their translation
property.

$U1A$ spin liquid:

\begin{eqnarray}
G_x(i) &=&e^{i\theta _x\tau ^3},  \nonumber \\
G_y(i) &=&e^{i\theta _y\tau ^3}, \\
u_{i,i+m} &=&i\rho _me^{i\phi _m\tau ^3}.  \nonumber
\end{eqnarray}

$U1B$ spin liquid:

\begin{eqnarray}
G_x(i) &=&e^{i(i_y\pi +\theta _x)\tau ^3},  \nonumber \\
G_y(i) &=&e^{i\theta _y\tau ^3}, \\
u_{i,i+m} &=&i\left( -\right) ^{m_yi_x}\rho _me^{i\phi _m\tau ^3}.  \nonumber
\end{eqnarray}

$U1C$ spin liquid:

\begin{eqnarray}
G_x(i) &=&e^{i\theta _x\tau ^3}(i\tau ^1),  \nonumber \\
G_y(i) &=&e^{i\theta _y\tau ^3}(i\tau ^1), \\
u_{i,i+m} &=&i\rho _me^{i(-)^i\phi _m\tau ^3}.  \nonumber
\end{eqnarray}

Next, we should find all the gauge inequivalent PSG's $G_{P_{xy}}\left(
i\right) $ and $G_{P_{x\bar{y}}}\left( i\right) $ for the above cases
respectively. In fact, as proved in Ref.\cite{Wqoslpub}, it exists the
following relations between $U1A$ and $U1B$ PSG's:

\begin{eqnarray}
\tilde{G}_{P_{xy}}(i) &=&\left( -\right) ^{i_xi_y}G_{P_{xy}}(i),  \nonumber
\\
\tilde{G}_{P_{x\bar{y}}}\left( i\right) &=&\left( -\right) ^{i_xi_y}G_{P_{x
\bar{y}}}\left( i\right) ,  \label{AB} \\
\tilde{G}_{T^{*}}(i) &=&G_{T^{*}}\left( i\right)  \nonumber
\end{eqnarray}
where $G_{P_{xy},P_{x\bar{y}}}$ are $U1A$ PSG's. Hence we can consider the $
U1A$ PSG's only, and using the relations, we can construct the $U1B$ PSG's.

1) $U1A$ spin liquid, $G_x(i)=e^{i\theta _x\tau ^3}$, $G_y(i)=e^{i\theta
_y\tau ^3}$, Eq.(\ref{PxyPxy}) reads 
\begin{eqnarray}
G_{P_{xy}}\left( i\right) &=&e^{i\left( \left( i_x-i_y\right) \varphi
_{pxy}+\phi _{pxy}\right) \tau ^3}  \label{GpxyA1} \\
\text{or }G_{P_{xy}}\left( i\right) &=&e^{i\left( \left( i_x+i_y\right)
\varphi _{pxy}+\phi _{pxy}\right) \tau ^3}\left( i\tau ^1\right) ,
\label{GPxyA2}
\end{eqnarray}
and 
\begin{eqnarray*}
G_{P_{x\bar{y}}}\left( i\right) &=&e^{i\left( \left( i_x+i_y\right) \varphi
_{px\bar{y}}+\phi _{px\bar{y}}\right) \tau ^3} \\
\text{or }G_{P_{x\bar{y}}}\left( i\right) &=&e^{i\left( \left(
i_x-i_y\right) \varphi _{px\bar{y}}+\phi _{px\bar{y}}\right) \tau ^3}\left(
i\tau ^1\right)
\end{eqnarray*}

Gauge transformation $W(i)=e^{-i\left( i_x\mp i_y\right) \varphi _{pxy}\tau
^3/2}$ transfers (\ref{GpxyA1}) and (\ref{GPxyA2}) into 
\begin{equation}
G_{P_{xy}}\left( i\right) =e^{i\phi _{pxy}\tau ^3}\text{ or }
G_{P_{xy}}\left( i\right) =e^{i\phi _{pxy}\tau ^3}\left( i\tau ^1\right) ,
\label{Gpxy}
\end{equation}
and do not change the forms of $G_{x,y,P_{x\bar{y}}}$. From the last
equation of (\ref{PxyPxy}), we obtain the following four cases,

a) 
\begin{eqnarray*}
G_{P_{xy}}\left( i\right) &=&e^{i\phi _{pxy}\tau ^3}, \\
G_{P_{x\bar{y}}}\left( i\right) &=&e^{i\left( \left( i_x+i_y\right) \varphi
_{px\bar{y}}+\phi _{px\bar{y}}\right) \tau ^3}
\end{eqnarray*}
Gauge transformation $W(i)=e^{-i\left( i_x+i_y\right) \varphi _{px\bar{y}
}/2\tau ^3}$ transfers the above into 
\begin{eqnarray}
G_{P_{xy}}\left( i\right) &=&e^{i\phi _{pxy}\tau ^3},  \nonumber \\
G_{P_{x\bar{y}}}\left( i\right) &=&e^{i\phi _{px\bar{y}}\tau ^3}
\label{U1Aa}
\end{eqnarray}
and does not change the forms of $G_{x,y}$.

b)

\begin{eqnarray}
G_{P_{xy}}\left( i\right) &=&e^{i\phi _{pxy}\tau ^3},  \nonumber \\
G_{P_{x\bar{y}}}\left( i\right) &=&\eta _{px\bar{y}}^ie^{i\phi _{px\bar{y}
}\tau ^3}\left( i\tau ^1\right) ,\eta _{px\bar{y}}=\pm 1  \label{U1Ab}
\end{eqnarray}

c)

\begin{eqnarray}
G_{P_{xy}}\left( i\right) &=&e^{i\phi _{pxy}\tau ^3}\left( i\tau ^1\right) ,
\nonumber \\
G_{P_{x\bar{y}}}\left( i\right) &=&\eta _{px\bar{y}}^ie^{i\phi _{px\bar{y}
}\tau ^3},\eta _{px\bar{y}}=\pm 1  \label{U1Ac}
\end{eqnarray}

d)

\begin{eqnarray*}
G_{P_{xy}}\left( i\right) &=&e^{i\phi _{pxy}\tau ^3}\left( i\tau ^1\right) ,
\\
G_{P_{x\bar{y}}}\left( i\right) &=&e^{i\left( \left( i_x-i_y\right) \varphi
_{px\bar{y}}+\phi _{px\bar{y}}\right) \tau ^3}\left( i\tau ^1\right)
\end{eqnarray*}
Gauge transformation $W(i)=e^{-i\left( i_x-i_y\right) \varphi _{px\bar{y}
}/2\tau ^3}$ transfers the above into 
\begin{eqnarray}
G_{P_{xy}}\left( i\right) &=&e^{i\phi _{pxy}\tau ^3}\left( i\tau ^1\right) ,
\nonumber \\
G_{P_{x\bar{y}}}\left( i\right) &=&e^{i\phi _{px\bar{y}}\tau ^3}\left( i\tau
^1\right)  \label{U1Ad}
\end{eqnarray}
and does not change the forms of $G_{x,y}$.

Now we add the last symmetry spin parity $T^{*}$. For the same reason, $
G_{T^{*}}\left( i\right) $ can take one of the two forms (\ref{U1Gauge}) and
(\ref{U1Gauge2}). The relations (\ref{PxyT*Txy}) require that 
\begin{eqnarray}
G_{T^{*}}\left( i\right) G_x\left( i\right) G_{T^{*}}^{-1}\left( i-\hat{x}
\right) G_x^{-1}\left( i\right) &=&e^{i\theta _{xt}\tau ^3}  \nonumber \\
G_{T^{*}}\left( i\right) G_y\left( i\right) G_{T^{*}}^{-1}(i-\hat{y}
)G_y^{-1}\left( i\right) &=&e^{i\theta _{yt}\tau ^3}  \nonumber \\
G_{T^{*}}\left( i\right) G_{P_{xy}}\left( i\right) G_{T^{*}}^{-1}\left(
P_{xy}\left( i\right) \right) G_{P_{xy}}^{-1}\left( i\right) &=&e^{i\theta
_{tpxy}\tau ^3}  \nonumber \\
G_{T^{*}}\left( i\right) G_{P_{x\bar{y}}}\left( i\right)
G_{T^{*}}^{-1}\left( P_{x\bar{y}}\left( i\right) \right) G_{P_{x\bar{y}
}}^{-1}\left( i\right) &=&e^{i\theta _{tpx\bar{y}}\tau ^3}  \nonumber \\
G_{T^{*}}^2\left( i\right) &=&e^{i\theta \tau ^3}.  \nonumber \\
\end{eqnarray}
Hence we have that all the $U1A$ type PSG's:

\begin{eqnarray}
G_{P_{xy}}\left( i\right) &=&e^{i\phi _{pxy}\tau ^3}  \nonumber \\
G_{P_{x\bar{y}}}\left( i\right) &=&e^{i\phi _{px\bar{y}}\tau ^3} \\
G_{T^{*}}\left( i\right) &=&\eta _t^ie^{i\phi _t\tau ^3}|_{\eta _t=-1},\eta
_t^ie^{i\phi _t\tau ^3}\left( i\tau ^1\right)  \nonumber
\end{eqnarray}

\begin{eqnarray}
G_{P_{xy}}\left( i\right) &=&e^{i\phi _{pxy}\tau ^3}  \nonumber \\
G_{P_{x\bar{y}}}\left( i\right) &=&\eta _{px\bar{y}}^ie^{i\phi _{px\bar{y}
}\tau ^3}\left( i\tau ^1\right) \\
G_{T^{*}}\left( i\right) &=&\eta _t^ie^{i\phi _t\tau ^3}|_{\eta
_t=-1},e^{i\phi _t\tau ^3}\left( i\tau ^1\right)  \nonumber
\end{eqnarray}

\begin{eqnarray}
G_{P_{xy}}\left( i\right) &=&e^{i\phi _{pxy}\tau ^3}\left( i\tau ^1\right) 
\nonumber \\
G_{P_{x\bar{y}}}\left( i\right) &=&\eta _{px\bar{y}}^ie^{i\phi _{px\bar{y}
}\tau ^3} \\
G_{T^{*}}\left( i\right) &=&\eta _t^ie^{i\phi _t\tau ^3}|_{\eta
_t=-1},e^{i\phi _t\tau ^3}\left( i\tau ^1\right)  \nonumber
\end{eqnarray}

\begin{eqnarray}
G_{P_{xy}}\left( i\right) &=&e^{i\phi _{pxy}\tau ^3}\left( i\tau ^1\right) 
\nonumber \\
G_{P_{x\bar{y}}}\left( i\right) &=&e^{i\phi _{px\bar{y}}\tau ^3}\left( i\tau
^1\right) \\
G_{T^{*}}\left( i\right) &=&\eta _t^ie^{i\phi _t\tau ^3}|_{\eta _t=-1},\eta
_t^ie^{i\phi _t\tau ^3}\left( i\tau ^1\right)  \nonumber
\end{eqnarray}

We can label these spin liquids as $U1Ag_{P_{xy}}(g_{P_{x\bar{y}}})_{\eta
_{px\bar{y}}}(g_{T^{*}})_{\eta _t}$, where $g_{P_{xy},P_{x\bar{y}}}=\tau
^{0,1}$ corresponding to $e^{i\phi _{pxy,px\bar{y}}\tau ^3}$ and $e^{i\phi
_{pxy,px\bar{y}}\tau ^3}\left( i\tau ^1\right) $. Now we consider the form
of ansatz that in invariant under the above PSG's. The translation symmetry
requires that

\begin{equation}
u_{i,i+m}=u_m=u_m^0\tau ^0+u_m^3\tau ^3.
\end{equation}

The 180$^{\circ }$ rotation symmetry requires that for $g_{P_{xy}}(g_{P_{x
\bar{y}}})_{\eta _{px\bar{y}}}=\tau ^0\tau _{+}^0,\tau ^1\tau _{+}^1,$

\begin{equation}
u_m=u_{-m}=u_m^{\dagger }
\end{equation}
and for $g_{P_{xy}}(g_{P_{x\bar{y}}})_{\eta _{px\bar{y}}}=\tau ^0\tau _{\eta
_{px\bar{y}}}^1,\tau ^1\tau _{\eta _{px\bar{y}}}^0,$

\begin{equation}
u_m=\eta _{px\bar{y}}^m\tau ^1u_{-m}\tau ^1=-\eta _{px\bar{y}}^mu_m
\end{equation}

The spin parity symmetry $T^{*}$ requires that for $(g_{T^{*}})_{\eta
_t}=\tau _{-}^0$

\begin{equation}
u_m=-\left( -\right) ^mu_m
\end{equation}
and for $(g_{T^{*}})_{\eta _t}=\tau _{\eta _t}^1$

\begin{equation}
u_m=-\eta _t^mu_m^0\tau ^0+\eta _t^mu_m^3\tau ^3
\end{equation}

We find that the following 6 sets of ansatz that give rise to $U(1)$
symmetric spin liquids:

$U1A[\tau ^0\tau _{-}^1,\tau ^1\tau _{-}^0]\tau _{-}^0$ 
\begin{eqnarray}
u_{i,i+m} &=&u_m^0\tau ^0+u_m^3\tau ^3  \nonumber \\
u_m^{0,3} &=&0,\text{ if }m=even
\end{eqnarray}
Here we have used the notation $U1A[ab,cd]e$ to represent the collection $
U1Aabe$ and $U1Acde$.

$U1A[\tau ^0\tau _{+}^0,\tau ^1\tau _{+}^1]\tau _{+}^1$ 
\begin{equation}
u_{i,i+m}=u_m^3\tau ^3
\end{equation}

$U1A[\tau ^0\tau _{+}^0,\tau ^1\tau _{+}^1]\tau _{-}^1$ 
\begin{eqnarray}
u_{i,i+m} &=&u_m^3\tau ^3  \nonumber \\
u_m^3 &=&0,\text{ if }m=odd
\end{eqnarray}
The ansatz gives rise to $U(1)\times U(1)$ spin liquids since $u_{ij}$ only
connect points within two different sublattices.

Other PSG's lead to vanishing ansatz or $SU(2)$ spin liquids and can be
dropped.

2) $U1B$ spin liquid, $G_x(i)=\left( -\right) ^{i_y}e^{i\theta _x\tau
^3},G_y(i)=e^{i\theta _y\tau ^3}.$ From relations (\ref{AB}), one can obtain
all the $U1B$ PSG's easily, 
\begin{eqnarray}
G_{P_{xy}}\left( i\right) &=&\left( -\right) ^{i_xi_y}e^{i\phi _{pxy}\tau
^3},  \nonumber \\
G_{P_{x\bar{y}}}\left( i\right) &=&\left( -\right) ^{i_xi_y}e^{i\phi _{px
\bar{y}}\tau ^3}, \\
G_{T^{*}}\left( i\right) &=&\eta _t^ie^{i\phi _t\tau ^3}|_{\eta _t=-1},\eta
_t^ie^{i\phi _t\tau ^3}\left( i\tau ^1\right)  \nonumber
\end{eqnarray}

\begin{eqnarray}
G_{P_{xy}}\left( i\right) &=&\left( -\right) ^{i_xi_y}e^{i\phi _{pxy}\tau ^3}
\nonumber \\
G_{P_{x\bar{y}}}\left( i\right) &=&\left( -\right) ^{i_xi_y}\eta _{px\bar{y}
}^ie^{i\phi _{px\bar{y}}\tau ^3}\left( i\tau ^1\right) \\
G_{T^{*}}\left( i\right) &=&\eta _t^ie^{i\phi _t\tau ^3}|_{\eta
_t=-1},e^{i\phi _t\tau ^3}\left( i\tau ^1\right)  \nonumber
\end{eqnarray}

\begin{eqnarray}
G_{P_{xy}}\left( i\right) &=&\left( -\right) ^{i_xi_y}e^{i\phi _{pxy}\tau
^3}\left( i\tau ^1\right)  \nonumber \\
G_{P_{x\bar{y}}}\left( i\right) &=&\left( -\right) ^{i_xi_y}\eta _{px\bar{y}
}^ie^{i\phi _{px\bar{y}}\tau ^3} \\
G_{T^{*}}\left( i\right) &=&\eta _t^ie^{i\phi _t\tau ^3}|_{\eta
_t=-1},e^{i\phi _t\tau ^3}\left( i\tau ^1\right)  \nonumber
\end{eqnarray}

\begin{eqnarray}
G_{P_{xy}}\left( i\right) &=&\left( -\right) ^{i_xi_y}e^{i\phi _{pxy}\tau
^3}\left( i\tau ^1\right) ,  \nonumber \\
G_{P_{x\bar{y}}}\left( i\right) &=&\left( -\right) ^{i_xi_y}e^{i\phi _{px
\bar{y}}\tau ^3}\left( i\tau ^1\right) , \\
G_{T^{*}}\left( i\right) &=&\eta _t^ie^{i\phi _t\tau ^3}|_{\eta _t=-1},\eta
_t^ie^{i\phi _t\tau ^3}\left( i\tau ^1\right)  \nonumber
\end{eqnarray}

We can label these spin liquids as $U1Bg_{P_{xy}}(g_{P_{x\bar{y}}})_{\eta
_{px\bar{y}}}(g_{T^{*}})_{\eta _t}$. Now we consider the form of ansatz that
in invariant under the above PSG's. The translation symmetry requires that 
\begin{equation}
u_{i,i+m}=\left( -\right) ^{m_yi_x}u_m=\left( -\right) ^{m_yi_x}\left(
u_m^0\tau ^0+u_m^3\tau ^3\right) .
\end{equation}

The 180$^{\circ }$ rotation symmetry requires that for $g_{P_{xy}}(g_{P_{x
\bar{y}}})_{\eta _{px\bar{y}}}=\tau ^0\tau _{+}^0,\tau ^1\tau _{+}^1,$ 
\begin{equation}
u_m=\left( -\right) ^{m_xm_y}u_m^{\dagger }
\end{equation}
and for $g_{P_{xy}}(g_{P_{x\bar{y}}})_{\eta _{px\bar{y}}}=\tau ^0\tau _{\eta
_{px\bar{y}}}^1,\tau ^1\tau _{\eta _{px\bar{y}}}^0,$ 
\begin{equation}
u_m=-\left( -\right) ^{m_xm_y}\eta _{px\bar{y}}^mu_m
\end{equation}

The spin parity symmetry $T^{*}$ requires that for $(g_{T^{*}})_{\eta
_t}=\tau _{-}^0$ 
\begin{equation}
u_m=-\left( -\right) ^mu_m
\end{equation}
and for $(g_{T^{*}})_{\eta _t}=\tau _{\eta _t}^1$ 
\begin{equation}
u_m=-\eta _t^mu_m^0\tau ^0+\eta _t^mu_m^3\tau ^3
\end{equation}

We find that the following 10 sets of ansatz that give rise to $U(1)$
symmetric spin liquids:

$U1B[\tau ^0\tau _{-}^1,\tau ^1\tau _{-}^0]\tau _{-}^0$ 
\begin{eqnarray}
u_{i,i+m} &=&\left( -\right) ^{m_yi_x}\left( u_m^0\tau ^0+u_m^3\tau ^3\right)
\nonumber \\
u_m^{0,3} &=&0,\text{ if }m=even
\end{eqnarray}

$U1B[\tau ^0\tau _{+}^0,\tau ^1\tau _{+}^1]\tau _{+}^1$ 
\begin{eqnarray}
u_{i,i+m} &=&\left( -\right) ^{m_yi_x}u_m^3\tau ^3  \nonumber \\
u_m^3 &=&0,\text{ if }m_x=odd\text{ and }m_y=odd
\end{eqnarray}

$U1B[\tau ^0\tau _{-}^1,\tau ^1\tau _{-}^0]\tau _{+}^1$ 
\begin{eqnarray}
u_{i,i+m} &=&\left( -\right) ^{m_yi_x}u_m^3\tau ^3  \nonumber \\
u_m^3 &=&0,\text{ if }m_x=even\text{ and }m_y=even
\end{eqnarray}

$U1B[\tau ^0\tau _{+}^0,\tau ^1\tau _{+}^1]\tau _{-}^1$ 
\begin{eqnarray}
u_{i,i+m} &=&\left( -\right) ^{m_yi_x}u_m^3\tau ^3  \nonumber \\
u_m^3 &=&0,\text{ if }m_x=odd\text{ or }m_y=odd
\end{eqnarray}
The ansatz gives rise to $(U(1))^4$ spin liquids since $u_{ij}$ only connect
points within four different sublattices.

$U1B[\tau ^0\tau _{-}^1,\tau ^1\tau _{-}^0]\tau _{-}^1$ 
\begin{eqnarray}
u_{i,i+m} &=&\left( -\right) ^{m_yi_x}\left( u_m^0\tau ^0+u_m^3\tau ^3\right)
\nonumber \\
u_m^0 &=&0,\text{ if }m=even  \nonumber \\
u_m^3 &=&0,\text{ if }m_x=even\text{ or }m_y=even
\end{eqnarray}

Other PSG's lead to vanishing ansatz or $SU(2)$ spin liquids and can be
dropped.

3) $U1C$ spin liquid, $G_x(i)=e^{i\theta _x\tau ^3}(i\tau ^1)$ and $
G_y(i)=e^{i\theta _y\tau ^3}(i\tau ^1)$. 
\[
G_{P_{xy}}\left( i\right) =e^{i\theta _{pxy}(i)\tau ^3},e^{i\theta
_{pxy}(i)\tau ^3}\left( i\tau ^1\right) . 
\]
For the former four equations in (\ref{PxyPxy}), we have that 
\begin{eqnarray*}
\theta _{pxy}(i) &=&\left( -\right) ^i\varphi _{pxy}+\phi _{pxy}, \\
&&i_x\pi +\left( -\right) ^i\varphi _{pxy}+\phi _{pxy}
\end{eqnarray*}
and the last three equations in (\ref{PxyPxy}) require that 
\begin{eqnarray}
G_{P_{xy}}\left( i\right) &=&\eta _{pxy}^i\eta _{xpxy}^{i_x}e^{i\left( \eta
_{xpxy}^i\pi /4+\phi _{pxy}\right) \tau ^3},  \nonumber \\
&&e^{i\left( \left( -\right) ^i\varphi _{pxy}+\phi _{pxy}\right) \tau
^3}\left( i\tau ^1\right)
\end{eqnarray}
and 
\begin{eqnarray}
G_{P_{x\bar{y}}}\left( i\right) &=&\eta _{px\bar{y}}^i\eta _{xpx\bar{y}
}^{i_x}e^{i\left( \eta _{xpx\bar{y}}^i\pi /4+\phi _{px\bar{y}}\right) \tau
^3},  \nonumber \\
&&e^{i\left( \left( -\right) ^i\varphi _{px\bar{y}}+\phi _{px\bar{y}}\right)
\tau ^3}\left( i\tau ^1\right)
\end{eqnarray}
where $\eta _{pxy,xpxy}=\pm 1,\eta _{px\bar{y},xpx\bar{y}}=\pm 1$. Now we
add the last symmetry spin parity $T^{*}$. In a similar way, through the
relations (\ref{PxyT*Txy}), we have that 
\begin{eqnarray}
G_{T^{*}}\left( i\right) &=&\eta _t^ie^{i\phi _t\tau ^3},  \nonumber \\
&&\eta _{xt}^{i_x}e^{i\left( \left( -\right) ^i\varphi _t+\phi _t\right)
\tau ^3}\left( i\tau ^1\right)
\end{eqnarray}
where $\eta _t=\pm 1,\eta _{xt}=\pm 1$.

It is noted that gauge transformation $W_i=e^{i(-)^i\phi \tau ^3}$ and $
W_i=\eta _{pxy}^{i_x}$ do not change $G_{x,y}$, and can be used to simplify
the forms of $G_{P_{xy},P_{x\bar{y}},T^{*}}$. We conclude all the $U1C$ type
PSG's: 
\begin{eqnarray}
G_{P_{xy}}\left( i\right) &=&\eta _{xpxy}^{i_x}e^{i\left( \eta _{xpxy}^i\pi
/4+\phi _{pxy}\right) \tau ^3}  \nonumber \\
G_{P_{x\bar{y}}}\left( i\right) &=&\eta _{px\bar{y}}^i\eta
_{xpxy}^{i_x}e^{i\left( \eta _{xpxy}^i\pi /4+\phi _{px\bar{y}}\right) \tau
^3} \\
G_{T^{*}}\left( i\right) &=&\eta _t^ie^{i\phi _t\tau ^3}|_{\eta _t=-1},\eta
_{xpxy}^{i_x}e^{i\phi _t\tau ^3}\left( i\tau ^1\right)  \nonumber
\end{eqnarray}

\begin{eqnarray}
G_{P_{xy}}\left( i\right) &=&\eta _{xpxy}^{i_x}e^{i\left( \eta _{xpxy}^i\pi
/4+\phi _{pxy}\right) \tau ^3}  \nonumber \\
G_{P_{x\bar{y}}}\left( i\right) &=&\eta _{px\bar{y}}^ie^{i\left( \eta
_{xpxy}^i\pi /4+\phi _{px\bar{y}}\right) \tau ^3}\left( i\tau ^1\right) \\
G_{T^{*}}\left( i\right) &=&\eta _t^ie^{i\phi _t\tau ^3}|_{\eta _t=-1},\eta
_{xpxy}^{i_x}e^{i\phi _t\tau ^3}\left( i\tau ^1\right)  \nonumber
\end{eqnarray}

\begin{eqnarray}
G_{P_{xy}}\left( i\right) &=&e^{i\left( \eta _{xpxy}^i\pi /4+\phi
_{pxy}\right) \tau ^3}\left( i\tau ^1\right)  \nonumber \\
G_{P_{x\bar{y}}}\left( i\right) &=&\eta _{px\bar{y}}^i\eta
_{xpxy}^{i_x}e^{i\left( \eta _{xpxy}^i\pi /4+\phi _{px\bar{y}}\right) \tau
^3} \\
G_{T^{*}}\left( i\right) &=&\eta _t^ie^{i\phi _t\tau ^3}|_{\eta _t=-1},\eta
_{xpxy}^{i_x}e^{i\phi _t\tau ^3}\left( i\tau ^1\right)  \nonumber
\end{eqnarray}

\begin{eqnarray}
G_{P_{xy}}\left( i\right) &=&e^{i\left( \eta _{xpxy}^i\pi /4+\phi
_{pxy}\right) \tau ^3}\left( i\tau ^1\right)  \nonumber \\
G_{P_{x\bar{y}}}\left( i\right) &=&\eta _{px\bar{y}}^ie^{i\left( \eta
_{xpxy}^i\pi /4+\phi _{px\bar{y}}\right) \tau ^3}\left( i\tau ^1\right) \\
G_{T^{*}}\left( i\right) &=&\eta _t^ie^{i\phi _t\tau ^3}|_{\eta _t=-1},\eta
_{xpxy}^{i_x}e^{i\phi _t\tau ^3}\left( i\tau ^1\right)  \nonumber
\end{eqnarray}

We can label these spin liquids as $U1C(g_{P_{xy}})_{\eta _{xpxy}}(g_{P_{x
\bar{y}}})_{\eta _{px\bar{y}}}(g_{T^{*}})$. Now we consider the form of
ansatz that in invariant under the above PSG's. The translation symmetry
requires that 
\begin{equation}
u_{i,i+m}=u_m^0\tau ^0+\left( -\right) ^iu_m^3\tau ^3.
\end{equation}

The 180$^{\circ }$ rotation symmetry requires that for $G_{P_{xy}}\left(
i\right) G_{P_{x\bar{y}}}(P_{xy}(i))=\eta _{px\bar{y}}^i\tau ^0$ 
\begin{equation}
u_m^0=-\eta _{px\bar{y}}^mu_m^0,u_m^3=(-\eta _{px\bar{y}})^mu_m^3;
\end{equation}
for $G_{P_{xy}}\left( i\right) G_{P_{x\bar{y}}}(P_{xy}(i))=\eta _{px\bar{y}
}^i\eta _{xpxy}^{i_y}\left( i\tau ^1\right) $

\begin{eqnarray}
u_m^0 &=&-\eta _{px\bar{y}}^m\eta _{xpxy}^{m_y}u_m^0,  \nonumber \\
u_m^3 &=&-(-\eta _{px\bar{y}})^m\eta _{xpxy}^{m_y}u_m^3.
\end{eqnarray}

The spin parity symmetry $T^{*}$ requires that for $G_{T^{*}}\left( i\right)
=\eta _t^ie^{i\phi _t\tau ^3}|_{\eta _t=-1}$ 
\begin{equation}
u_{i,i+m}=-\left( -\right) ^mu_{i,i+m}
\end{equation}
and for $G_{T^{*}}\left( i\right) =\eta _{xpxy}^{i_x}e^{i\phi _t\tau
^3}\left( i\tau ^1\right) $ 
\begin{equation}
u_m^0=-\eta _{xpxy}^{m_x}u_m^0,u_m^3=\eta _{xpxy}^{m_x}u_m^3
\end{equation}

We find that the following 14 sets of ansatz that give rise to $U(1)$
symmetric spin liquids:

$U1C[\tau _{+}^0\tau _{-}^0,\tau _{-}^0\tau _{-}^0,\tau _{+}^1\tau
_{-}^1,\tau _{-}^1\tau _{-}^1]\tau ^0$ 
\begin{eqnarray}
u_{i,i+m} &=&u_m^0\tau ^0+\left( -\right) ^iu_m^3\tau ^3  \nonumber \\
u_m^{0,3} &=&0,\text{ if }m=even;
\end{eqnarray}

$U1C[\tau _{+}^0\tau _{+}^0,\tau _{+}^1\tau _{+}^1]\tau ^1$ 
\begin{eqnarray}
u_{i,i+m} &=&\left( -\right) ^iu_m^3\tau ^3  \nonumber \\
u_m^3 &=&0,\text{ if }m=odd
\end{eqnarray}
The ansatz gives rise to $U(1)\times U(1)$ spin liquids.

$U1C[\tau _{+}^0\tau _{-}^0,\tau _{+}^1\tau _{-}^1]\tau ^1$ 
\begin{equation}
u_{i,i+m}=\left( -\right) ^iu_m^3\tau ^3;
\end{equation}

$U1C[\tau_{-}^0\tau _{+}^0,\tau _{-}^1\tau _{+}^1]\tau ^1$ 
\begin{eqnarray}
u_{i,i+m} &=&\left( -\right) ^iu_m^3\tau ^3  \nonumber \\
u_m^3 &=&0,\text{ if }m_x=odd\text{ or }m_y=odd;
\end{eqnarray}
The ansatz gives rise to $(U(1))^4$ spin liquids.

$U1C[\tau _{-}^0\tau _{-}^0,\tau _{-}^1\tau _{-}^1]\tau ^1$ 
\begin{eqnarray}
u_{i,i+m} &=&u_m^0\tau ^0+\left( -\right) ^iu_m^3\tau ^3  \nonumber \\
u_m^0 &=&0,\text{ if }m_x=even\text{ or }m_y=odd  \nonumber \\
u_m^3 &=&0,\text{ if }m_x=odd;
\end{eqnarray}

$U1C[\tau _{-}^0\tau _{-}^1,\tau _{-}^1\tau _{-}^0]\tau ^1$ 
\begin{eqnarray}
u_{i,i+m} &=&u_m^0\tau ^0+\left( -\right) ^iu_m^3\tau ^3  \nonumber \\
u_m^0 &=&0,\text{ if }m_x=even  \nonumber \\
u_m^3 &=&0,\text{ if }m_x=odd\text{ or }m_y=even;
\end{eqnarray}

Other PSG's lead to vanishing ansatz or $SU(2)$ spin liquids and can be
dropped.

\section{Classification of $SU(2)$ Spin Liquids}

We assume that for a $SU(2)$ spin liquid, one can always choose a gauge so
that $u_{ij}$ has a form 
\begin{equation}
u_{ij}=u_{ij}^0\tau ^0
\end{equation}
Hereafter we call this gauge canonical gauge. In the canonical gauge, IGG
has a form $\mathcal{G}=SU(2)$. The gauge transformations that keep $u_{ij}$
to have the form in the canonical gauge are given by 
\begin{equation}
W_i=\eta (i)g
\end{equation}
where $\eta (i)=\pm 1$ for each $i$ and $g\in SU(2).$ The gauge
transformation $G_{x,y}$ associated with the translation also take the above
form 
\begin{eqnarray}
G_x(i) &=&\eta _x(i)g_x  \nonumber \\
G_y(i) &=&\eta _y(i)g_y
\end{eqnarray}
And we can gauge fix the above form as 
\begin{eqnarray}
G_x(i) &=&g_x  \nonumber \\
G_y(i) &=&\eta _y(i)g_y
\end{eqnarray}
where $\eta _y(i_x=0,i_y)=1.$ Now the relation (\ref{TxTy}) reads 
\begin{equation}
\eta _y(i-\hat{x})\eta _y(i)\in SU(2)
\end{equation}
We find that there are only two different PSG's for translation symmetric
ansatz 
\begin{eqnarray}
G_x(i) &=&g_x,G_y(i)=g_y \\
\text{and }G_x(i) &=&g_x,G_y(i)=\left( -\right) ^{i_x}g_y
\end{eqnarray}
The two PSG's lead to the following two translation symmetric ansatz 
\begin{eqnarray}
u_{i,i+m} &=&u_m^0\tau ^0 \\
\text{and }u_{i,i+m} &=&\left( -\right) ^{m_xi_y}u_m^0\tau ^0
\end{eqnarray}
Similar to $Z_2$ case, adding more symmetries, we have all the $SU(2)$ PSG's
belonged to triangular lattices.

$SU2A\eta _{px\bar{y}}\eta _t$
\begin{eqnarray}
G_x(i) &=&g_x,  \nonumber \\
G_y(i) &=&g_y,  \nonumber \\
G_{T^{*}}(i) &=&\eta _t^ig_{T^{*}}, \\
G_{P_{xy}}(i) &=&g_{P_{xy}},  \nonumber \\
G_{P_{x\bar{y}}}(i) &=&\eta _{px\bar{y}}^ig_{P_{x\bar{y}}},  \nonumber
\end{eqnarray}
and

$SU2B\eta _{px\bar{y}}\eta _t$
\begin{eqnarray}
G_x(i) &=&g_x,  \nonumber \\
G_y(i) &=&\left( -\right) ^{i_x}g_y,  \nonumber \\
G_{T^{*}}(i) &=&\eta _t^ig_{T^{*}}, \\
G_{P_{xy}}(i) &=&\left( -\right) ^{i_xi_y}g_{P_{xy}},  \nonumber \\
G_{P_{x\bar{y}}}(i) &=&\left( -\right) ^{i_xi_y}\eta _{px\bar{y}}^ig_{P_{x
\bar{y}}}  \nonumber
\end{eqnarray}
where $g_x,g_y,g_{T^{*}},g_{P_{xy}},g_{P_{x\bar{y}}}\in SU(2),\eta _t=\pm 1,$
and $\eta _{px\bar{y}}=\pm 1$. Moreover, only $\eta _t=-1$ and $\eta _{px
\bar{y}}=-1$ lead to non vanishing ansatz. So there are only two kinds of $
SU(2)$ spin liquids,

$SU2A$ spin liquids: 
\begin{equation}
u_{i,i+m}=u_m^0\tau ^0,m=odd\text{,}  \label{SU2A}
\end{equation}

$SU2B$ spin liquids: 
\begin{equation}
u_{i,i+m}=\left( -\right) ^{m_xi_y}u_m^0\tau ^0,m=odd\text{.}  \label{SU2B}
\end{equation}

\bibliographystyle{apsrev}
\bibliography{/home/wen/bib/wencross,/home/wen/bib/publst,/home/wen/bib/all}

\end{document}